\documentclass{aa}  

\usepackage{graphicx}
\usepackage[varg]{txfonts}
\usepackage{lipsum}
\usepackage{subcaption}     
\usepackage{lscape}         
\usepackage{placeins}       
\usepackage{tabularx}
\usepackage{longtable}
\usepackage{adjustbox}
\usepackage{array} 
\usepackage{booktabs}
\usepackage{geometry}

\usepackage{hyperref}
\hypersetup{
    colorlinks=true,         
    linkcolor=blue,          
    citecolor=blue,          
    filecolor=blue,          
    urlcolor=blue            
}
\usepackage{natbib}
\bibpunct{(}{)}{;}{a}{}{,}

\usepackage{soul}
\usepackage{cancel}
\usepackage{ulem}

\begin{document}

\title{Computing Natural Magnitudes in the Photometric Systems of Astronomical Plates using \textit{Gaia} DR3 SEDs}

   \titlerunning{Computing Natural Magnitudes in Plate Photometric Systems using \textit{Gaia} SEDs}
   
   \subtitle{}

\author{Maryam Raouph
   \and  Andreas Schrimpf  }

\institute{History of Astronomy and Observational Astronomy, Physics Department, Philipps University Marburg, Germany\\
    \email{maryam.raouph@physik.uni-marburg.de}
     \email{andreas.schrimpf@physik.uni-marburg.de}
}

\date{Accepted on 29 December 2025}

\abstract
{Accurate photometric calibration of astronomical photographic plates remains a fundamental challenge in astronomy, especially when bridging historical photographic data with modern observations due to the mismatch of spectral sensitivities of photographic plates and passbands of modern calibration catalogs.}
{We intend to derive consistent natural magnitudes for celestial sources within the intrinsic photometric systems of astronomical photographic plates by using \textit{Gaia} Data Release 3 (DR3) blue photometer (BP) and red photometer (RP) low-resolution spectral data and to show its superiority to former methods.}
{We compiled spectral characteristic data for emulsions and filters applied in photometric observations using glass plates. The collected color sensitivities, modified by atmospheric reddening depending on the air mass, are then used to compute accurate natural magnitudes and fluxes of objects in the photographic plates through synthetic photometry, utilizing a catalog of \textit{Gaia} spectral energy distributions (SEDs) over the range 330 nm $\leq$ $\lambda$ $\leq$ 1050 nm (XP spectra). This process uses GaiaXPy, a Python library designed to handle \textit{Gaia} DR3 spectral data. These natural magnitudes are then compared with results from the color term method used to compile the data in existing photoplate archives.}
{Comparing the synthetic magnitudes with those existing in the Archives of Photographic PLates for Astronomical USE (APPLAUSE), we were able to reveal systematic errors of the existing data in the range of $\pm 0.3$ mag and higher. 
In addition, the presented method allows for an accommodation of stars with similar color index but of different luminosity classes as well as an effective correction of atmospheric reddening at higher air masses, approximately 0.2 mag.}
{}

\keywords{Methods: data analysis -- Astronomical data bases -- Virtual observatory tools -- Techniques: photometric
               }

\maketitle

\section{Introduction}
\label{sec:intro}
Astronomical photographic plates, often referred to simply as 'plates,' have historically been fundamental tools in advancing observational astronomy. Despite the meticulous efforts of traditional astronomers, their recordings suffered from inherent inaccuracies and subjectivities present in handwritten descriptions, drawings, and measurements. However, the advent of plates in the late 19th century revolutionized the field by enabling the capture of objective images of celestial objects, free from the biases and limitations of human perception.

After a few different developments, the glass plate with a dry light-sensitive emulsion prevailed. Edward Charles Pickering introduced the professional use of photographic plates for scientific applications in astronomy at the Harvard College Observatory (HCO) in 1882. Max Wolf began to exploit the new technique at the Heidelberg Observatory in 1891. Many observatories were soon convinced of the great benefits of astronomical photographic plates as detectors and data storage devices \citep{Schrimpf2024}.

Photographic plates have been used in astronomy for more than a century. With more than 70 different plate collections worldwide, there are estimated to be over eight million of these invaluable resources \citep{Hudec2019}. The HCO houses the world's largest and oldest astronomical photographic glass plate collection in the northern and southern hemispheres, with more than 550\,000 pieces, some dating back 144 years \citep{Grindlay2009}. The scanned collection is available through a virtual library known as Digital Access to a Sky Century at Harvard (DASCH) \citep{Laycock2010}. The Sonneberg Observatory in Thuringia, Germany, holds the second largest collection of patrol plates, covering the entire northern sky and comprising around 300\,000 plates taken between 1923 and 2010 \citep{Kroll2009}. Most of these plates have been digitized; however, they are not available via any online service.
The APPLAUSE database (Archives of Photographic PLates for Astronomical USE\footnote{\url{https://www.plate-archive.org/}})  currently has about 70\,000 digitized direct photographic plates (among others), mostly from the German observatories Bamberg, Potsdam, Hamburg, and Tautenburg, as well as from Tartu (Estonia), and the Vatican Observatory \citep{Enke2024}. Almost the entire Chinese stock of astronomical plates, roughly 30\,000, has been digitized and partly analyzed. Data related to these analyzed plates, as well as scanned images, are available via a web service of the Chinese National Astronomical Data Center \citep{Shang2024}. The French NAROO project was initiated to digitize roughly about 100\,000 plates from French observatories \citep{Robert2021}.

Photographic plates typically range in size from $6 \times 6 \, 
\mathrm{cm}^2$ to $30 \times 30 \, \mathrm{cm}^2$ \citep{Kroll2019}, although the largest ever used measured about $1 \times 1 \, \mathrm{m}^2$, held at Yerkes Observatory\footnote{International Glass Plates Group (IGPG) \url{https://osf.io/chj43/wiki/Historical Research Using Photographic Plates}} \citep{Osborn2024}. Plates have been used in surveys, ranging from high-resolutions of 1-2 $\arcsec$ to low-resolution campaigns of about $15 \arcsec$ \citep{Jones1971}. High-resolution plates most often utilize blue-sensitive emulsions (pg = photographic), while lower-resolution surveys may employ a combination of blue (pg) and yellow to red (pv = photo visible) sensitive emulsions \citep{Braeuer1999}, often supplemented with additional filters. Moreover, the composition of emulsions has evolved over time, further complicating the spectral response. While blue emulsions generally exhibit relatively consistent sensitivity, yellow emulsions vary significantly. 

\begin{figure*}[h!]
	\centering
	\includegraphics[width = 0.495\hsize]{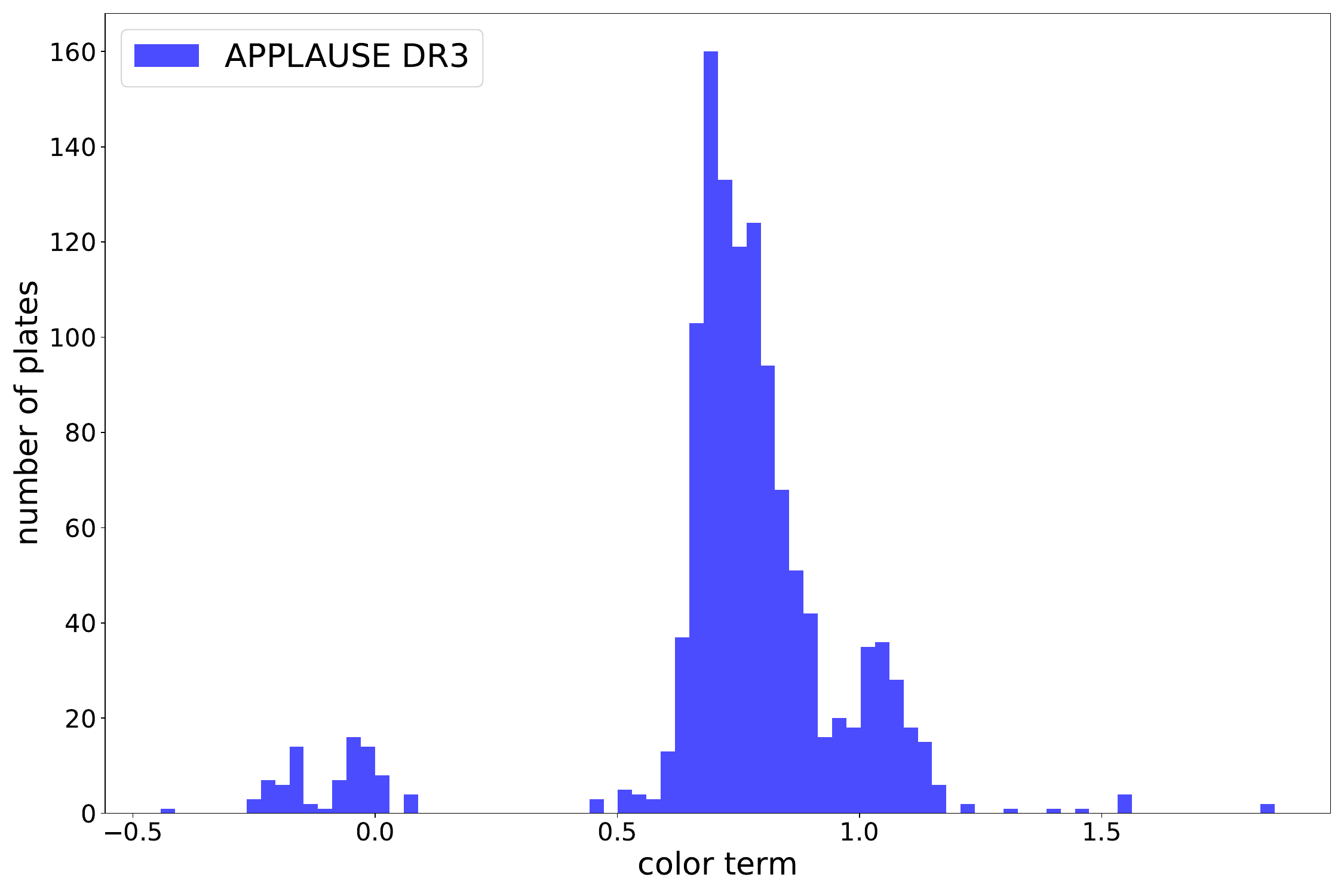}
	\includegraphics[width = 0.495\hsize]{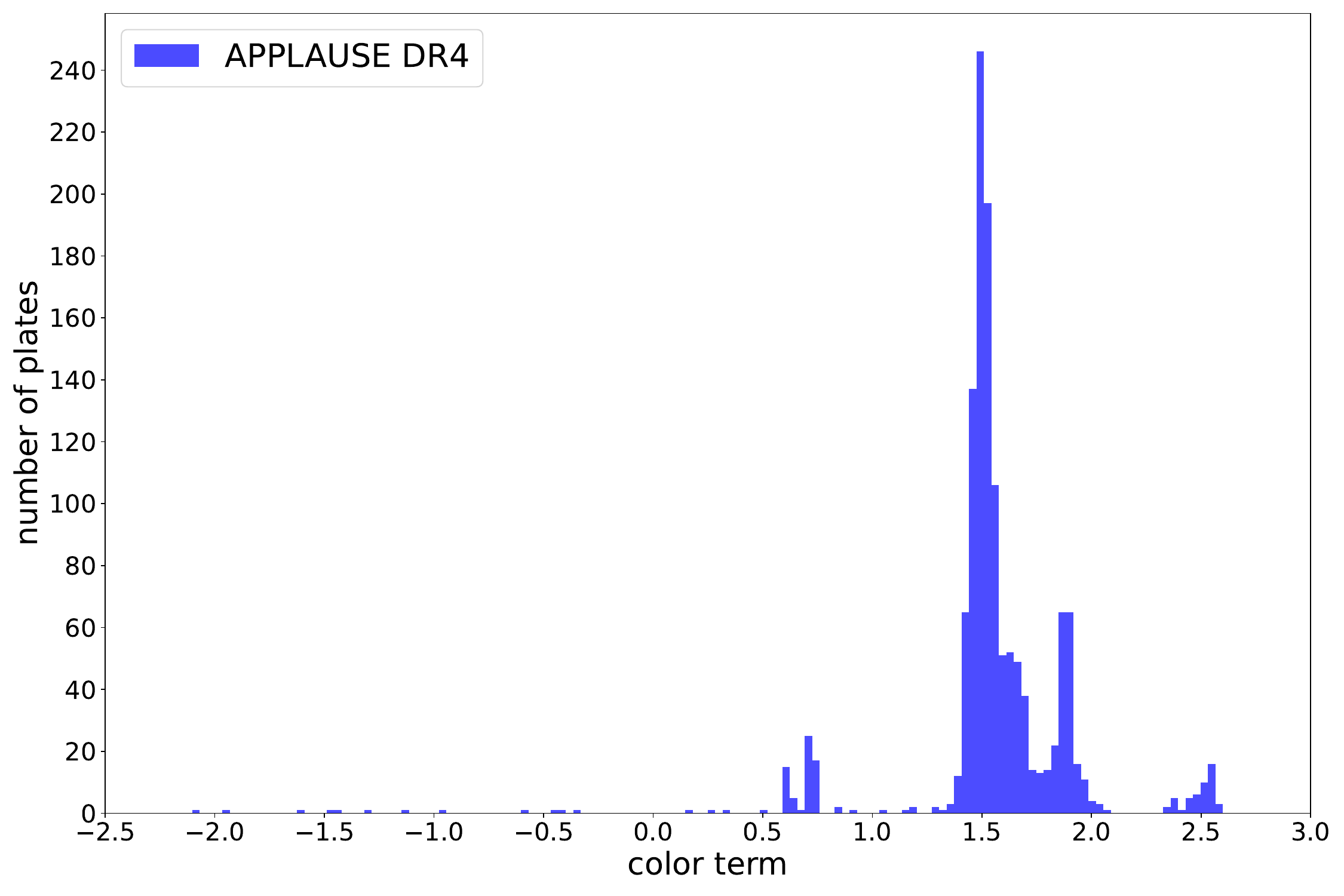}
	\caption{The color terms derived from a series of plates from two data releases of APPLAUSE archive for the randomly selected constant star UCAC4 600-107181. Left: Data release 3 using the Tycho-2 and APASS DR9 as a reference catalog. Right: Data release 4 using the \textit{Gaia} EDR3 as a reference catalog.}
	\label{fig1}
\end{figure*}

The typical mean error in the magnitudes determined from sources in photographic plates is in the range of 0.15 to 0.3 mag. This is mainly a statistical error due to the bad quality of photographic plates as detectors in comparison with modern CCD or CMOS detectors.
A significant challenge with using photo plates as detectors is their spectral response. Before the development of star catalogs with common color systems, e.g., the Johnson \citep{johnson1963} or the Sloan Digital Sky Survey (SDSS) \citep{fukugita1996} photometric systems, the brightness of stars could only be compared within series of observations, which used identical emulsions. When different catalogs with sufficient sky coverage became available, empirical color transformations were established (see for example \cite{jordi2006}) that were and still are used to convert all data to a common photometric system necessary for an astrophysical interpretation of the observations. Similar transformations have also been used for compiling the data of the photoplate archives DASCH and APPLAUSE. However, such transformations are valid only within certain limits, and it is well accepted that they suffer from systematic errors, e.g., offsets in zero points.

With the availability of low-resolution spectra of about 200 million stars from the latest \textit{Gaia} data release, synthetic photometry can supply absolute photometry in any photometric system within the range of the spectra \citep{Montegriffo2023a} and thus avoid systematic errors.

In this paper, we present a study that uses synthetic photometry from Gaia low-resolution spectra to calculate the magnitude of calibration stars in the natural color system of the photographic plates and compares the results with those from existing data in archives.
In Section \ref{sec:color_term}, the former method to convert the color systems of reference catalogs to the color response of plate emulsions is introduced. 
Our solution, utilizing the \textit{Gaia} low-resolution BP/RP spectra, is explained in Section \ref{sec:SED}. 
A comparison of catalog data from the APPLAUSE archive and our results using \textit{Gaia} SEDs, revealing the systematic errors of the empirical color transformations, is presented in Section \ref{sec:comparison}, and Section \ref{sec:newapps} covers further corrections as well as a new application of synthetic photometry in the analysis of photoplates. The final section comprises a discussion of the results and outlook.

\section{Natural magnitudes using color systems of reference catalogs}
\label{sec:color_term}

Examining digitized astronomical plates necessitates addressing two primary inquiries: identifying sources through their coordinates, known as astrometry, and gauging the brightness of these sources, termed photometry. However, analyzing the plates poses more challenges than examining CCD images. This is largely attributed to the extensive field of view (FOV) typically covered by astronomical photo plates, exceeding that of CCDs. Consequently, devising a comprehensive plate solution for an entire photo plate proves challenging due to the inherent difficulties in achieving acceptable accuracy across such vast FOVs. Another notable difficulty arises from the intrinsic characteristics of the emulsion coating the plates. Unlike digital sensors, which typically exhibit a linear response to light, the emulsion coating on the plates often possesses a logarithmic intensity characteristic \citep{Martin1976, Spasovic2016}. This non-linearity complicates converting raw pixel values into physical units and efforts to accurately quantify and correct for variations in brightness across the plate. The use of different emulsions over time causes the plates to have non-uniform color sensitivity. Furthermore, the challenges of severe vignetting and optical aberrations, such as distortion, and the absence of flat exposures in the plates present an obstacle to precise calibration efforts.

However, the DASCH project presented an astrometric and photometric analysis pipeline that effectively addressed geometric distortion, vignetting, emulsion sensitivity variations, and atmospheric effects. \cite{Laycock2010} utilized a higher-order polynomial distortion to calculate plate solving, dividing the plate into concentric regions and calibrating the intensity for each region separately. Moreover, the APPLAUSE archive introduced PyPlate \citep{Tuvikene2014, Tuvikene2019}, a Python-based open-source software for processing digital images of direct astronomical photo plates. Tuvikene et al. enhanced the astrometric plate solution of flatbed-scanned plates by incorporating corrections for non-uniform scanner arm movements. Additionally, they adopted Laycock et al.'s approach to calibrate concentric regions for the photometric solution.

\subsection{The color term approach}
A significant challenge with using photo plates as detectors is their spectral response. Applying a polynomial correction based on the color index of the sources is a common approach for taking care of the mismatch of the photometric systems of the plates and reference catalogs, i.e., converting between the two differing photometric systems. This correction accounts for various influencing factors, including the wavelength-dependent transmission properties of the atmosphere, telescope, camera optics, and emulsion response.  However, experience shows that higher-order polynomial corrections do not improve accuracy or reduce errors. Therefore, both \cite{Laycock2010} and \cite{Tuvikene2014} employ a linear correction approach in their respective studies\footnote{To our knowledge, the Chinese digitization project uses this approach, too, as was presented by H. Yuan on the PDPP workshop 2024 \citep{Yuan2024}}.

The magnitude of the sources in the natural photometric system of the plate, using a catalog with magnitudes given in the UBV-color system, is defined as:

\begin{equation}
  m_{\mathrm{cat}} = V_{\mathrm{cat}} + c (B_{\mathrm{cat}} - V_{\mathrm{cat}})  \,,
\label{eq1}
\end{equation}

where $m_{\mathrm{cat}}$ represents the magnitude of the stars in the reference catalog transformed into the plate system, $B_{\mathrm{cat}}$ and $V_{\mathrm{cat}}$ are the Johnson $B$ and $V$ magnitudes of the stars in the reference catalog, respectively, and $c$ is the color term of the plate. $m_{\mathrm{cat}}$ is what we call the natural magnitude of the calibration stars in the plate's color system\footnote{The above definition of the natural magnitude is used by APPLAUSE. DASCH reverts the color index in this equation, resulting in negative color terms.}.

It is worth mentioning that $m_{\mathrm{cat}}$ is used to calibrate the raw magnitudes of stars determined from the photo plates. However, this calibration step is not part of this paper; it's implemented and well-tested in the two pipelines used to create the DASCH (daschlab) and APPLAUSE (PyPlate) databases.

In both pipelines, the color term is determined by minimizing the root mean square (RMS) of the photometric calibration. A color term $c=0$ indicates that the plate's color response closely matches that of the Johnson $V$ passband, while $c=1$ signifies alignment with the Johnson $B$ passband. Values of $c$ outside the range of 0 to 1 suggest that the color system of the reference catalog does not adequately match the spectral sensitivity of the plates. Most common photometric analyses of photographic plates use reference catalogs with Johnson-Cousins passband; some studies may employ different color systems. e.g., APPLAUSE DR4 uses the \textit{Gaia} EDR3 $G_{\mathrm{BP}}$ and $G_{\mathrm{RP}}$ passband \citep{Enke2024}.

In Figure \ref{fig1}, we present color terms for a series of plates containing the randomly selected constant star UCAC4 \citep{Zacharias2013} 600-107181 from the two most recent data releases of the APPLAUSE archive (DR3, DR4). This star is cataloged in the \textit{Gaia} DR3 under the identifier 2030169850461844224 and also holds the number 2153-2118-1 in the Tycho-2 catalog \citep{Høg2000}. The images were captured using two distinct instruments: the Lippert-Astrograph and the large Schmidt-Mirror at Bergedorf Observatory in Hamburg \citep{Anderson2004}.  In Data Release 3, calibration was performed with the help of the Tycho-2 and UCAC4 catalogs for astrometry, and the APASS DR9 \citep{Henden2016} and Tycho-2 catalogs for photometry. In Data Release 4, the extracted sources were calibrated in photometry and astrometry using the \textit{Gaia} Early Data Release 3 (EDR3) catalog.

Most of the plate emulsions employed at Bergedorf are highly sensitive to blue light, with peaks around $0.7$ in the left plot of Fig.\ \ref{fig1}. These emulsions include Kodak 103a-O (blue), Kodak 103a-D (visible), Kodak IIa-O (almost blue), Kodak Oa-0 (almost blue), as well as Agfa Astro Z and V (almost blue). The right plot in Fig. \ref{fig1} (APPLAUSE DR4) exhibits a shifted range for the color term, which will be further discussed in the next subsection.

\subsection{The color term difficulties}
The color term concept has several limitations. Firstly, it simplifies the entire spectral sensitivity differences of the plates and the passband of the reference catalogs into a single numerical value, which appears to be a rather crude approximation. Secondly, determining color terms relies on fitting procedures, though it is not a physical measure of the data. In cases where the same emulsion is used in taking data, the detector's spectral response remains constant, obviating the need for fitting. Consequently, the distribution of color terms reflects not only the properties of the plates but also a combination of data errors and convolution between the spectral energy distribution of the sources and the plates' color response. 

\begin{figure}
   \centering
   \resizebox{\hsize}{!}{\includegraphics{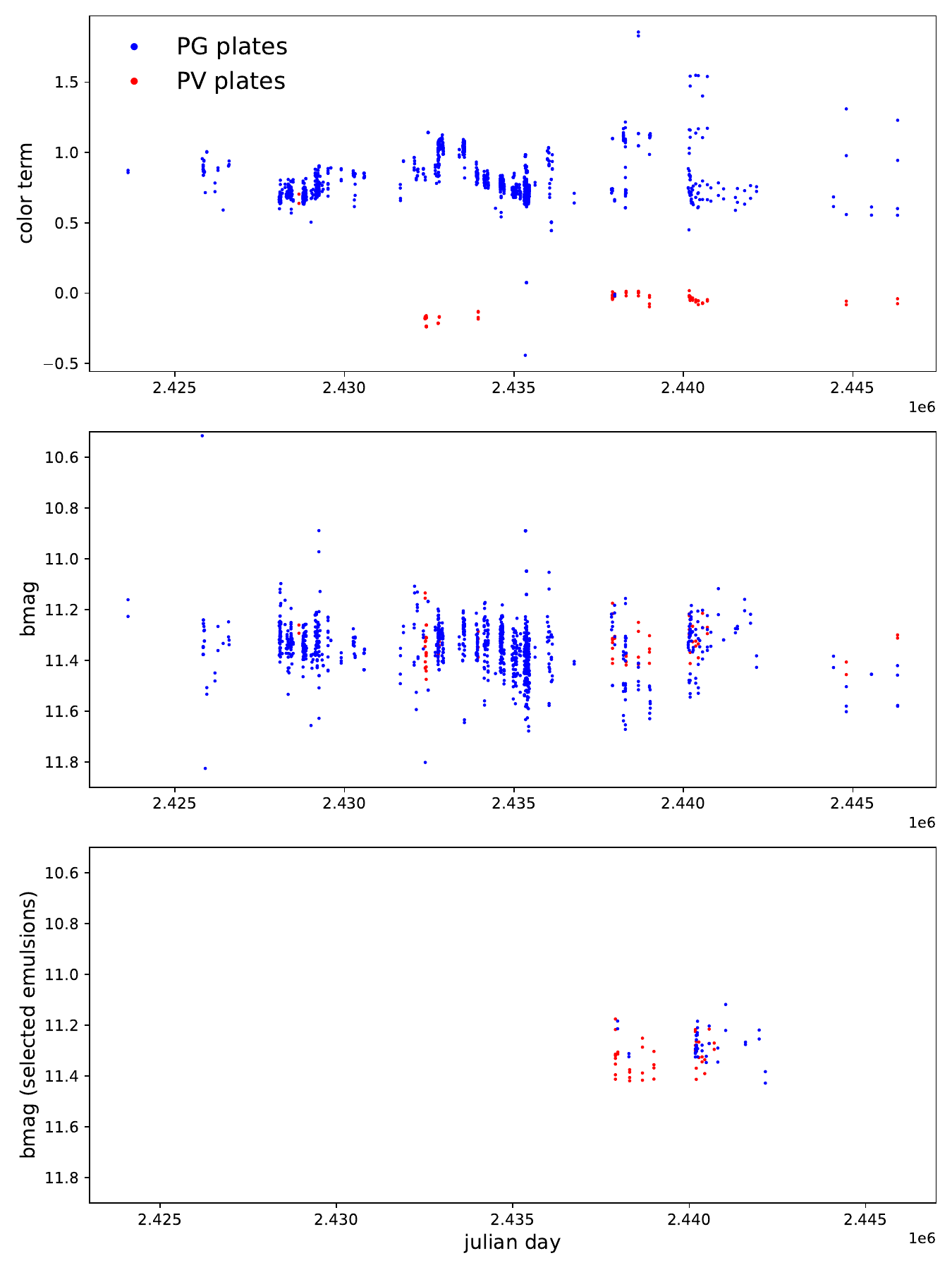}}
      \caption{Time series of the constant star UCAC4 600-107181 (color index $B-V = 0.39\ \mathrm{mag}$) from 1\,181 scans of 606 photographic plates, over 32 years, covering data from blue sensitive (pg, shown as blue dots) and yellow and red sensitive (pv, shown as red dots) plates from APPLAUSE DR3. Plates were classified based on known emulsions and reliable data (see appendix \ref{sec:appendix:A} for details about the plates used in this figure). Upper panel: Series of color terms determined by RMS fitting. Lower two panels: Calibrated $B$ magnitudes of the star, differing in that panel 3 includes  80 plates of the two emulsions Kodak IIa-O and Kodak 103a-D, exposed in the Lippert-Astrograph, only. The calibrated $V$ and $B$ magnitudes are identical except for a shift of the color index.}

         \label{fig2}
   \end{figure}

In Figure \ref{fig2}, a clear correlation is observed between the variations in color terms (upper panel) and the $B$ magnitudes of the constant star UCAC4 600-107181 (the two lower panels), with data derived from a series of APPLAUSE DR3 plates. Panels 1 and 2 include all emulsions used in the database, split into pv and pg categories, whereas panel 3 is restricted to two specific emulsions, Kodak IIa-O and Kodak 103a-D, exposed in the Lippert-Astrograph, only, relevant to the analysis described in Sect.\ \ref{sec:comparison:series}. The correlation here highlights a systematic error in the analysis. %
The natural magnitudes of one and the same star in a series of plates with varying emulsions reflect the distribution of color terms of the plates (see Eq.\ \ref{eq1}). The observed (instrumental) intensity of the red star UCAC4 600-107181, color index $B-V = 0.39\, \text{mag}$, yields lower brightness in a blue-sensitive plate compared to an observation with a visible-light-sensitive plate. However, the natural magnitude of the star in Fig. \ref{fig2}, determined by minimizing the RMS of the photometric calibration using the comparison catalog, does not perfectly match the variation of the observations in the plate series, leading to variations in the calibrated $B$ magnitude (Fig.\ \ref{fig2}). I.e., the natural magnitudes of this star in the APPLAUSE archive are erroneous, most probably because the fitted color terms are erroneous. Light curves from the DASCH archive reveal identical problems: e.g., the light curve of the star TYC 9504-35-1 with $B-V = 1.52\, \text{mag}$ from the DASCH archive shows a jump of about 2 mag in the calibrated magnitude from different plates\footnote{\url{https://dasch.cfa.harvard.edu/dr7/colorterms/}}.

DASCH and APPLAUSE (DR1 to DR3, PyPlate 3) utilize multiple catalogs, including UCAC4, APASS, KIC \citep{Brown2011}, and GSC \citep{Lasker2021}, for both astrometric and photometric calibration. While this approach reduces gaps in sky coverage, it introduces potential systematic errors. 

To address these challenges, the photometric calibration of APPLAUSE DR4, using the \textit{Gaia}, $G_{\mathrm{BP}}$ and $G_{\mathrm{RP}}$ passband, faces a significant issue: The color terms range from $-2$ to $2.5$ (see Fig. \ref{fig1}, right), indicating a large discrepancy in color response when compared to the \textit{Gaia} passband.

\begin{figure}[h!]
    \centering
    \resizebox{\hsize}{!}{\includegraphics{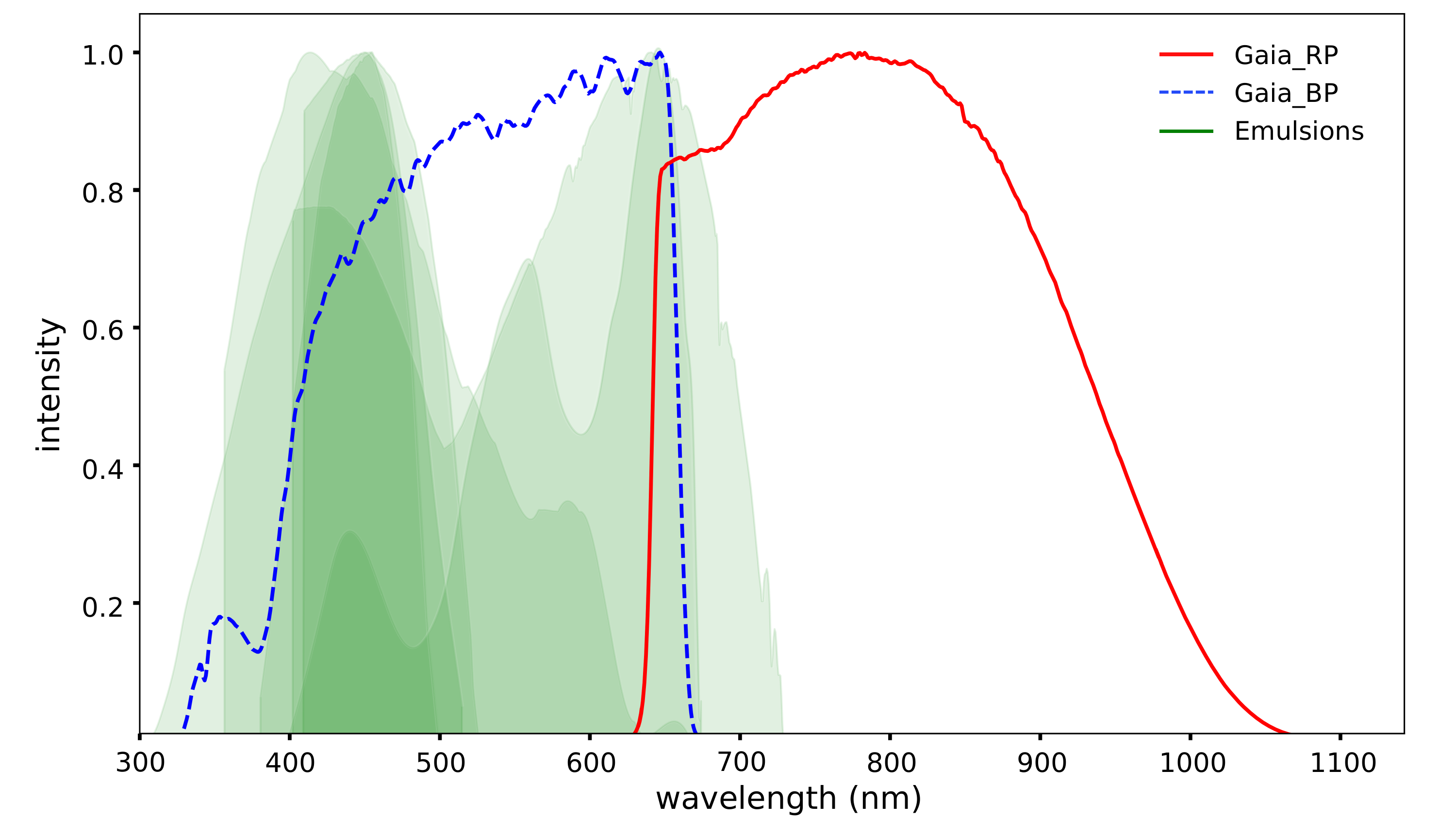}}
    \caption{Spectral sensitivities of various pg and pv plates coated with different emulsions are shown relative to the \textit{Gaia} photometric system. The green areas represent the plate's emulsion color responses, which already include the atmospheric transmission data, using the air mass 1.5 to 0 ratio, while the dashed blue and solid red lines indicate \textit{Gaia}'s blue and red photometer sensitivities, respectively.}
    \label{fig3}
\end{figure}

In Figure \ref{fig3}, the green shading illustrates the spectral sensitivities of various emulsions used for the plates across the \textit{Gaia} $G_{\mathrm{BP}}$ and $G_{\mathrm{RP}}$ passband. The plot reveals minor differences among the blue (pg) plates from Eastman Kodak 103a-O \citep{STScI_GSC_Surveys} and 103a-D \citep{Kodaknotebook1967}, and Orwo ZU21 and ZP1, and Agfa Astro and Astro Panchromatic 
\citep{Dokuchaeva1983}, shown in a darker green shade in the overlapping regions. However, significant variations exist in the spectral responses of the visible (pv) plates. Notably, while there is a good spectral overlap with the \textit{Gaia} $G_{\mathrm{BP}}$ passband, there is almost no spectral overlap with the \textit{Gaia} $G_{\mathrm{RP}}$ passband.
Consequently, the color correction is only marginally influenced by the \textit{Gaia} $G_{\mathrm{RP}}$ magnitudes, thus increasing the color term in the fitting procedure to numbers higher than one (see Fig.\ref{fig1}). Mainly, the blue passband of \textit{Gaia} is utilized in the color term concept, which does not align well, either, and is roughly a factor of 2 wider than the plates' sensitivities. Thus, using \textit{Gaia} $G_{\mathrm{BP}}$ and $G_{\mathrm{RP}}$ passbands does not seem to be very appropriate for photometric calibration of photo plates.

To tackle these issues, an advanced approach involves utilizing the BP/RP low-resolution spectral data from the \textit{Gaia} catalog, data release 3.

\begin{figure}[htp!]
	\centering
    \resizebox{\hsize}{!}{\includegraphics{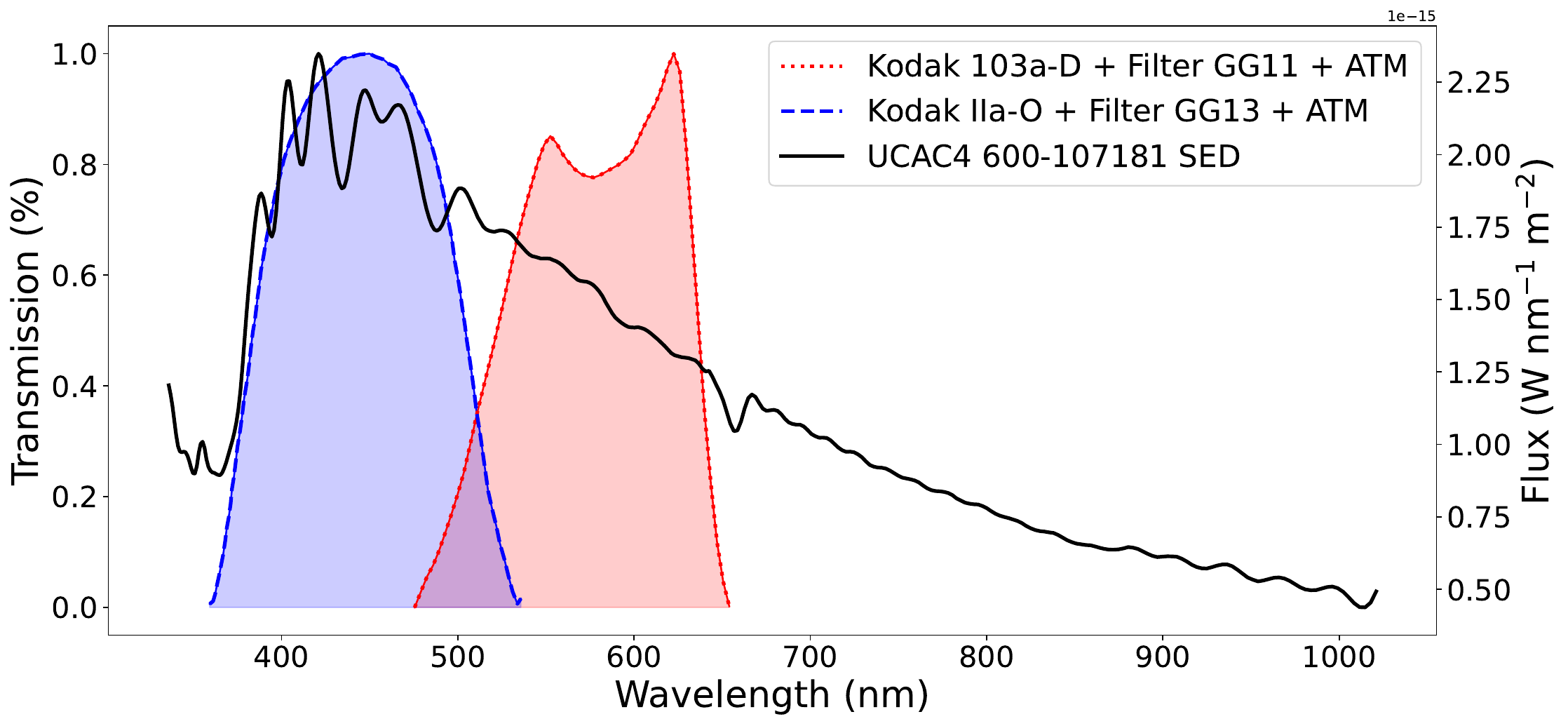}}
	\caption{The low-resolution spectrum of UCAC4 600-107181 is compared with two spectral sensitivity curves of Kodak emulsions with Schott filters for photographic photometry. The \textit{Gaia} SED is shown with a black solid line (right axis). The blue-filled plot with the dashed line represents the spectral sensitivity of the Kodak IIa-O emulsion combined with the transmission of a 2 $\mathrm{mm}$ Schott GG13 (=385) filter, which closely matches the Johnson-Cousins B band (blue-sensitive plate). The red-filled plot with dotted line illustrates Kodak 103a-D emulsion plus a 2 $\mathrm{mm}$ Schott GG11 filter, which closely matches the Johnson-Cousins V band (visible-light-sensitive plate). The influence of atmospheric transmission was taken into account for both spectral sensitivity curves at zenith (air mass 1). All spectral curves here have been normalized to their individual maximum.}
	\label{fig4}
\end{figure}

\section{Natural magnitudes using low-resolution spectra}
\label{sec:SED}
The innovative approach to more advanced photometric calibration in this study involves replacing the linear correction of the color term concept with the idea of using spectral energy distribution (SED) data from the latest \textit{Gaia} data release. 
The recently released \textit{Gaia} Data Release 3 \citep{Vallenari2023} includes flux-calibrated low-resolution spectra obtained from the BP and RP spectrophotometers in the wavelength range of 330 to 1050 {nm} (XP spectra) for more than 200 million sources, with magnitude G < 17.65. The availability of these externally calibrated (EC) XP spectra \citep{DeAngeli2023} enabled us to generate wide- and medium-band synthetic photometry to supply absolute optical photometry of stars in any passband fully enclosed in this wavelength range covered by \textit{Gaia} $G_{\mathrm{BP}}$ and $G_{\mathrm{RP}}$ spectra \citep{Montegriffo2023a}. Fig.\ \ref{fig4} shows the SED of our constant sample star from \textit{Gaia} DR3 combined with the spectral sensitivities of the plates used in our study.

To facilitate extracting \textit{Gaia} DR3 BP/RP mean spectra and calculating synthetic photometry, the free Python library called GaiaXPy\footnote{\url{https://gaia-dpci.github.io/GaiaXPy-website/}, version 2.1.2 for this study} has been released. 
Synthetic photometry is founded on a normalized mean flux \citep{Bessell2012}, which is obtained by integrating the product of the spectral sensitivity of the emulsion, including the transmission of any filter and the atmosphere $ S(\lambda) $, and spectral energy distribution over a given spectral range $f_\lambda (\lambda)$.
The mean flux in any photometric system can be expressed as \citep{Montegriffo2023a}:
\begin{equation}
 \displaystyle  	\left\langle f_\lambda \right\rangle = \frac{\int f_\lambda (\lambda) S(\lambda) \lambda\, d\lambda}{\int  S(\lambda) \lambda\, d\lambda}
 \, ,
\label{eq2}
\end{equation}

 where the factor $\lambda$ accounts for converting spectral energy to the number of photons detected \footnote{see for instance\\ \url{https://pysynphot.readthedocs.io/en/latest/properties.html\#bandpass-equivalent-monochromatic-flux}}.

%
The synthetic flux can be converted into a magnitude by
\begin{equation}
m_\mathrm{synth} = -2.5 \log \langle f_\lambda \rangle + \mathit{ZP}\, .
\label{eq3}
\end{equation}

\noindent
The computation of the zero point $ZP$ is based on a reference SED. We use the VEGAMAG magnitude system,  an unreddened A0V star with $V$ = 0.0 as a reference point. Then $ZP$ is calculated by
\begin{equation}
\mathit{ZP} = +2.5 \log \langle f_{\lambda} \rangle + V
\label{eq4}
\end{equation}
\noindent

The synthetic photometry utility of the GaiaXPy library uses the method \texttt{generate}, applying Eqs.\ \ref{eq2} to \ref{eq4} to return a dataframe with the generated synthetic photometry results. Magnitudes, fluxes, and flux errors are computed for each filter. The desired photometric system can be selected from a list of embedded systems or uploaded as a user-defined new system. User-defined photometric systems have to be converted to an internal representation similar to the \textit{Gaia} $G_{\mathrm{BP}}$ and $G_{\mathrm{RP}}$ spectra. This is kindly provided by the SVO Filter Profile Service \citep{svo2020}.

\section{Assessing the two methods}

\label{sec:comparison}

{Using the full spectral information of a star, the SED, to calculate the response in a color system of a photographic plate should yield more reliable results than applying a Taylor expansion of the color index. In order to check this out, we compared the natural magnitudes from the APPLAUSE archive as well as from synthetically Johnson-Cousins magnitudes from the Gaia DR3, with those determined by our approach. }

\subsection{The sample}

We selected photographic plates available in the APPLAUSE archive coated with Kodak IIa-O \citep{Anki2021} and Kodak 103a-D emulsions \citep{Moro2000}, widely used for stellar magnitude measurements. Kodak IIa-O emulsion is sensitive to the spectrum’s ultraviolet and blue regions, covering wavelengths from approximately 320 nm to 500 nm. On the other hand, Kodak 103a-D exhibits high sensitivity across the green to yellow regions of the visible spectrum, spanning 400 nm to 650 nm. To match the Johnson B band many of the blue emulsions have been used with a 2 mm Schott GG13 filter, which was renamed GG385 over the years, and for the yellow emulsions, a 2 mm Schott GG11 filter, later renamed GG495, was added \citep{noirlab_schott_filters}\footnote{\url{https://www.oapd.inaf.it/sede-di-asiago/telescopes-and-instrumentations/publications-and-archives/photographic-archive/plates-digitisation}}. Last but not least, the transmission of the atmosphere has to be considered; the plates were recorded at ground-based observatories. We corrected the spectral sensitivity of the plate and filter combinations with the ratio of the observation's air mass solar reference spectrum to the air mass 0, or extraterrestrial solar spectrum\footnote{\url{https://www.nrel.gov/grid/solar-resource/spectra-am1.5}} (see section \ref{sec:newapps:air_mass}). Some sensitivity functions were already calculated under Earth's atmospheric transmission, so no additional corrections were required \citep{Bell1972}. The final two spectral sensitivities used in this study are shown in Fig.\ \ref{fig4}. All parameters of the plates and sources necessary for our analysis, e.g., color term, UCAC4, Tycho2, and \textit{Gaia} DR3 IDs of stars, catalog magnitudes of calibration stars, color index of stars, natural magnitudes of catalog stars in the plates color system, can be retrieved via a Python API from the APPLAUSE archive. The DASCH archive, unfortunately, does not provide all of the data from their analysis, so we limited our test to analyzing plates from the APPLAUSE archive.

\subsection{Comparison with $m_\text{cat}$ from APPLAUSE}
\label{sec:comparison:color_index}

\begin{figure*}[htp!]
	\centering
	\includegraphics[angle=0,width=0.495\hsize]{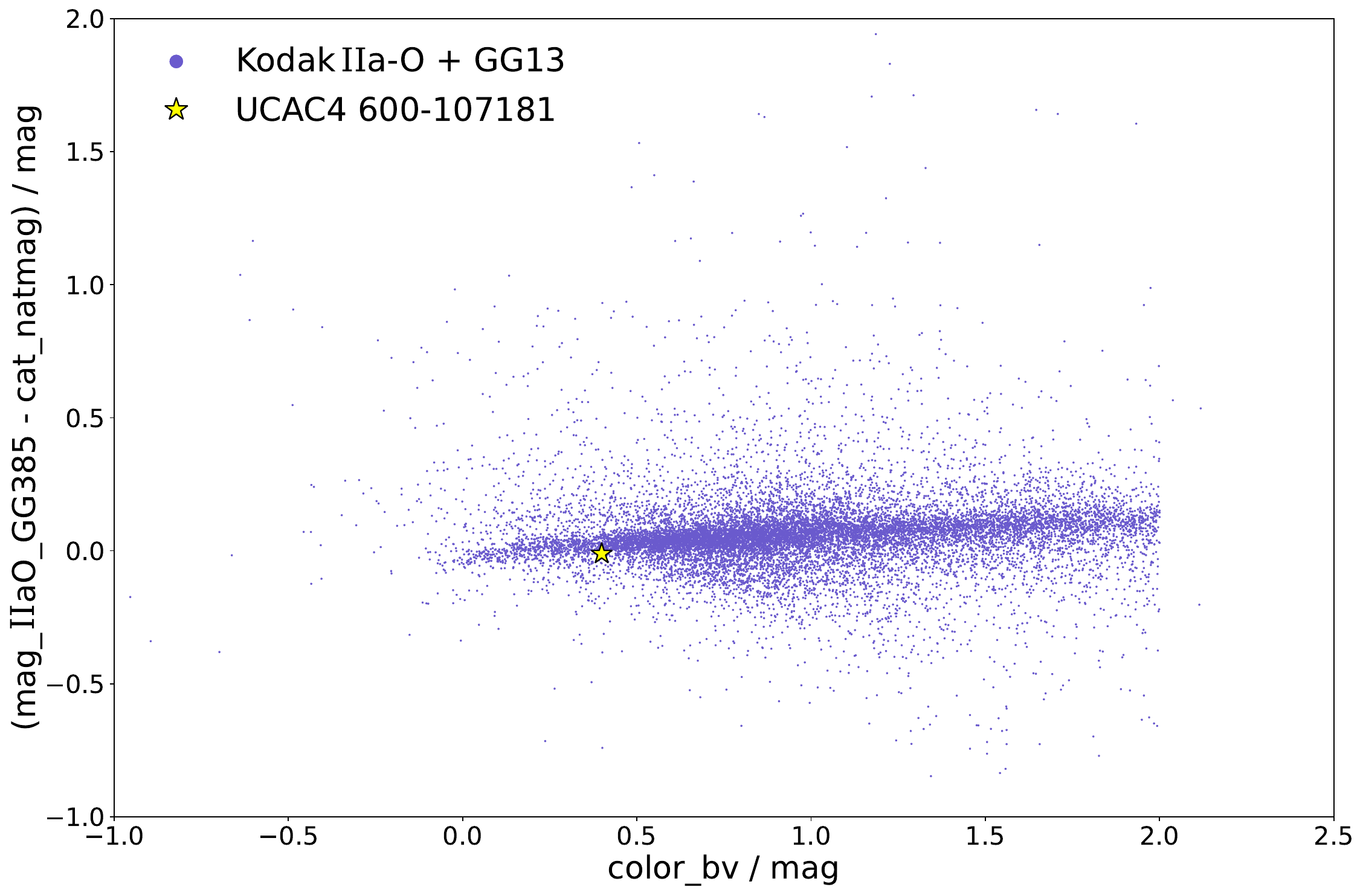}
	\includegraphics[angle=0,width=0.495\hsize]{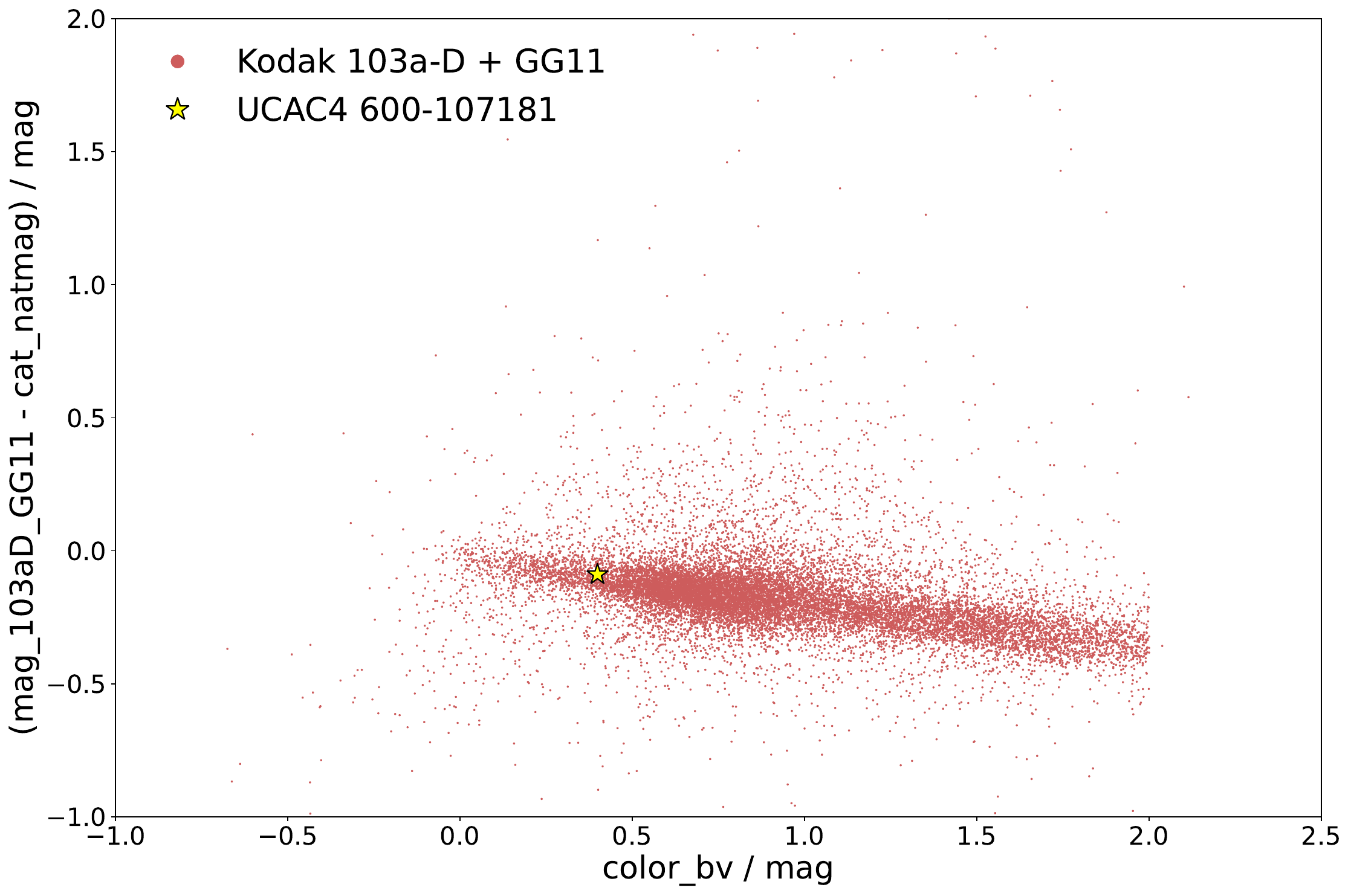}
	\caption{Difference $\Delta = \mathrm{m}_\mathrm{synth} -  m_\mathrm{cat}$ versus color index $\mathrm{color\_bv} = (B-V)$ for all in APPLAUSE DR3 identified, cleaned and with Gaia DR3 matched stars. Left: plate 12368, Großer Schmidt-Spiegel, Hamburg Observatory, 12.12.1968, Kodak IIa-O with filter GG13(=385), scan 14850, color term 0.83. Right: plate 11626, Großer Schmidt-Spiegel, Hamburg Observatory, 05.10.1964, Kodak 103a-D with filter GG11, scan 14057, color term 0.01. We used an air mass of 1.25 for the analysis in both cases. These plates in APPLAUSE have two scanned files with different color terms, producing two distributions. To keep this analysis figure simple, only one distribution for each such plate is shown.}
	\label{fig5}
\end{figure*}  

In the first test, we are addressing the dependence of the APPLAUSE $m_\mathrm{cat}$  on the color index $(B-V)$ in Eq.\ \ref{eq1}.

From our sample, we selected one of the blue plates and one of the visible plates each and analyzed the difference $\Delta = m_\mathrm{synth} -  m_\mathrm{cat}$ for all in APPLAUSE DR3 identified stars used for photometric calibration. We cleaned the data according to several quality flags available in APPLAUSE and matched them with Gaia DR3 sources (see appendix \ref{sec:appendix:B}).
These differences $\Delta$ versus the color index of the stars are shown in Fig.\ \ref{fig5}. 

In both cases, we find systematic variations in the differences with the color index of the stars. 
For the visible plate (right panel), there is an almost linear negative trend up to about $-0.3$ mag for the higher color indices, indicating that the determined color term is too large by $\approx 0.3/2 = 0.15$. For the blue plate (left panel), we observe a) a small increasing trend up to about $+0.15$ mag -- in this case, the color term is too small by $\approx 0.15/2 = 0.075$ -- and b) an increasing spread up to $\pm 0.7\ \mathrm{mag}$ for higher indices. The sharp cut-off at color index = 2 is due to the limit, which is used in PyPlate. It seems the authors were suspicious about using the color term fitting for higher color indices.

\subsection{Comparison with synthetic $m_\text{JKC}$ from Gaia DR3}
\label{sec:comparison:jkc}

\begin{figure*}[htp!]
	\centering
	\includegraphics[angle=0,width=0.495\hsize]{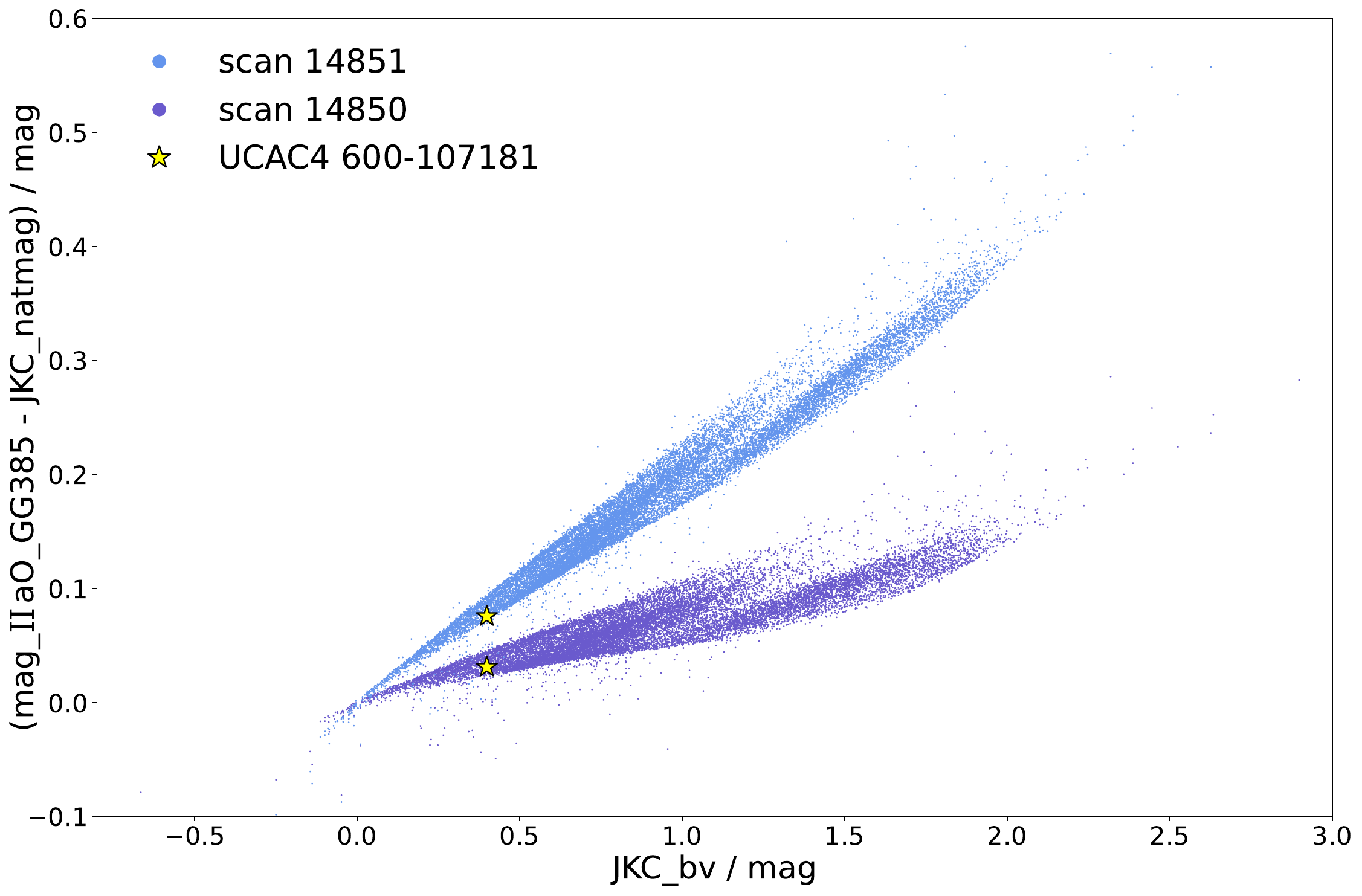}
	\includegraphics[angle=0,width=0.495\hsize]{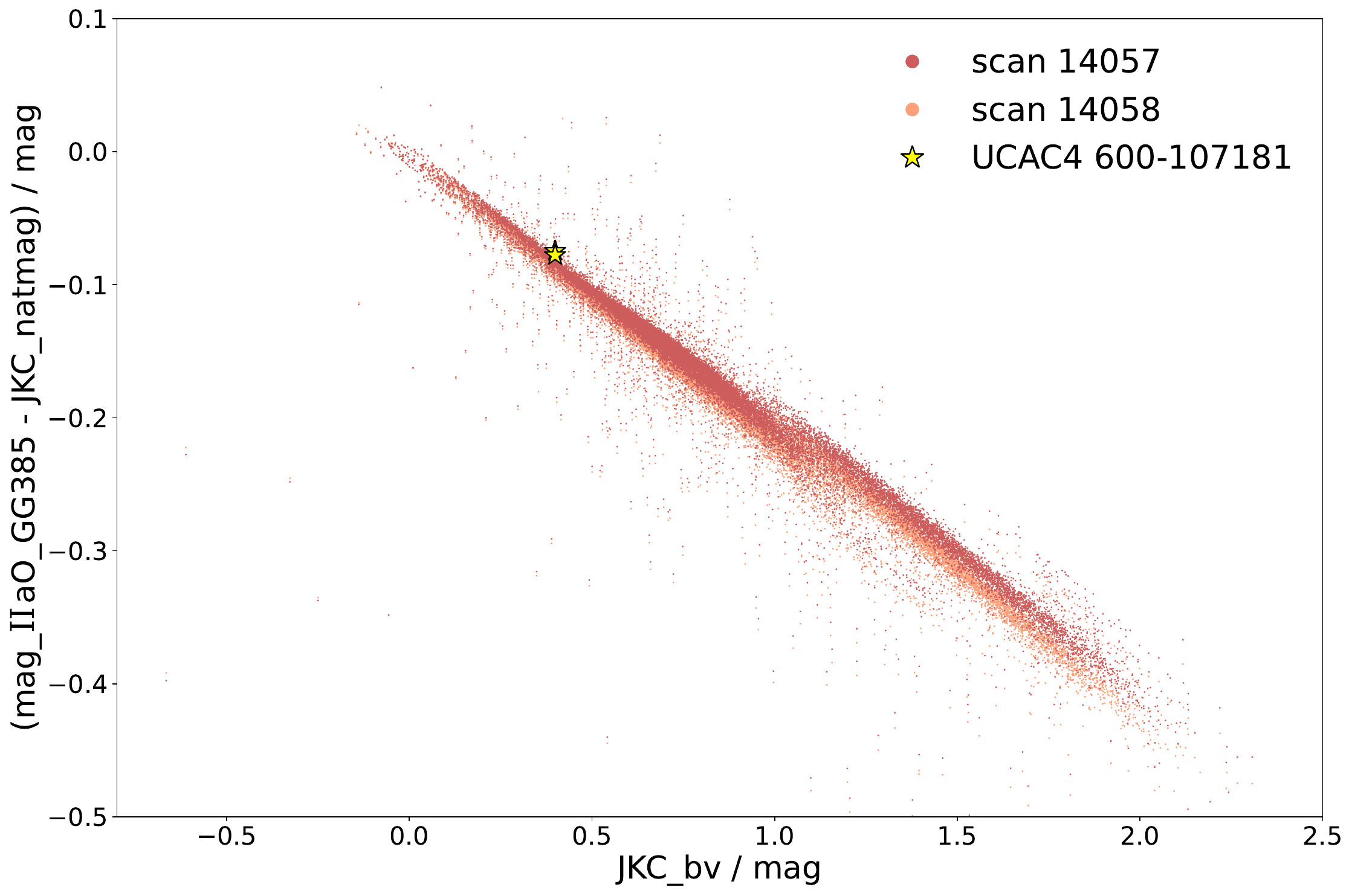}
	\caption{Difference $\Delta = \mathrm{m}_\mathrm{synth} - m_\mathrm{JKC}$ versus color index $\mathrm{JKC\_bv} = (B_\mathrm{JKC} - V_\mathrm{JKC}) $ for all in APPLAUSE DR3 identified, cleaned and with Gaia DR3 matched stars using synthetic Johnson-Cousins magnitudes from Gaia DR3. Left: plate 12368, Großer Schmidt-Spiegel, Hamburg Observatory, 12.12.1968, Kodak IIa-O with filter GG13(=385), scan 14850 (color term 0.826) and scan 14851 (color term 0.704). Right: plate 11626, Großer Schmidt-Spiegel, Hamburg Observatory, 05.10.1964, Kodak 103a-D with filter GG11, scan 14057 (color term 0.013), scan 14058 (color term 0.004). We used an air mass of 1.25 for the analysis in both cases. The ranges of the y axes were changed in comparison to Fig. \ref{fig5}.}
	\label{fig6}
\end{figure*}  

In the first test, we compared natural magnitudes APPLAUSE data determined from UCAC4 and APASS catalogs using color terms with those calculated from the Gaia SEDs of the stars. However, to compare the two methods, we need to use an identical data set in both approaches. Thus, we replaced the UCAC4 and APASS magnitudes with synthetic Johnson-Cousins magnitudes from Gaia DR3,

\begin{equation}
	m_{\mathrm{JKC}} = V_{\mathrm{JKV}} + c (B_{\mathrm{JKC}} - V_{\mathrm{JKC}})  \,,
	\label{eq5}
\end{equation}

where the index JKC marks the Gaia synthetic magnitudes. However, we did not modify the PyPlate code yet, so we could not recalculate the color terms according to Eq.\ \ref{eq5}. We used the APPLAUSE color terms instead. Fig.\ \ref{eq6} shows the differences $\Delta = m_\mathrm{synth} -  m_\mathrm{JKC}$ for all in APPLAUSE DR3 identified, cleaned, and with Gaia DR4 matched stars of the same two plates, which were used in Fig.\ \ref{fig5}. 

First, we find nearly identical general linear trends in the blue and visible plates. This time we plotted data for the two scans of each of the plates. Especially for the blue plate, the two scans differ significantly! We definitely can state that the color terms were not determined correctly. For a further analysis of the color term variations, see next subsection.

Second, the large spread shown in Fig.\ \ref{fig5} is no longer showing up in this analysis, and we find a few stars with color indices higher than 2. The obvious explanation of this finding is that the magnitude data from the UCAC4 and APASS catalog, which were used in APPLAUSE DR3 are erroneous, to a much greater extent than expected!

And third, Fig.\ \ref{fig6} reveals non-linear effects in the range of 0.1 mag, which show up when using higher quality data for the calibration stars.

The fine structure in Fig.\ \ref{fig6} will be discussed in Sect.\ \ref{sec:newapps:luminosity_classes}.

\subsection{Analyzing a series of plates}
\label{sec:comparison:series}

One of the results of the tests presented above is that the determination of the color terms in APPLAUSE is doubtful, which explains the effects that we found in Fig. \ref{fig2}, the variation of the ''constant'' magnitude as an effect of the variation of the color terms! To get a more qualitative view, we determined the color term errors by linear regression from differences, such as shown in Figs.\ref{fig5} and \ref{fig6}, using 40 plates from each color (blue and visible), including both scans, with identical emulsion-filter combinations exposed in one and the same telescope. The analysis revealed a mean offset in the color terms of 0.13 and a quite large spread of $0.06$ for the blue plates and -0.1 $\pm\ 0.02$ for visible plates.

For a star with a small color index of 0.39 mag (for example UCAC4 600-107181,  see Fig. \ref{fig2}) this results in a mean difference $\Delta = m_\mathrm{synth} -  m_\mathrm{cat}$ of 0.05 $\pm\ 0.02\ \mathrm{mag}$ in blue plates and -0.04 $\pm\ 0.01\ \mathrm{mag}$ in the visible plates, and for stars with a higher color index of 2 mag, an error of 0.26 $\pm\ 0.12\ \mathrm{mag}$ in the blue and -0.20 $\pm\ 0.04\ \mathrm{mag}$ in the visible plates. 
	
For the randomly chosen test star UCAC4 600-107181, the derived color term errors lead to apparent magnitude differences which are smaller than the listed $B$ magnitude variations of APPLAUSE shown in panel 3 of Fig.\ \ref{fig2}. This discrepancy may be accounted for by the errors of the calibration catalogs UCAC4 and APPAS as revealed in Sect.\ \ref{sec:comparison:jkc} or by a location of the star near the edges of the plates (in our sample, in 54 of 80 plates), which results in larger calibration errors. We need to check this when implementing our method into PyPlate and calculating calibrated magnitudes.

Besides the noticeable offsets of the color terms, the spread of the color term offsets in identical emulsions exposed in identical optics is about one magnitude larger than the error of the color terms given in the APPLAUSE database.

\section{Further modifications and applications}

\label{sec:newapps}

\subsection{Correcting effects of the air mass}
\label{sec:newapps:air_mass}

For ground-based astronomical observations, the atmosphere acts as a filter, leading to an extinction and a reddening. The effects of the atmosphere change with the air mass, which can be described as follows \citep{stetson2013}:
\begin{equation}
	m_{\mathrm{instr}} = m_{\mathrm{obs}} - \kappa({\lambda}) X \,,
	\label{eq6}
\end{equation}
where the observed raw magnitude $m_{\mathrm{obs}}$ has to be reduced by the product of the wavelength-dependent extinction coefficient $\kappa({\lambda})$ and the air mass $X$ to get the instrumental magnitude, later used for calibration. The air mass is usually approximated with $X \approx \sec z$, with the zenith distance $z$. With this approximation, we get $X \approx 1.15$ at a zenith distance of $30^\circ$, $\approx 1.41$ at $45^\circ$, and $\approx 2$ at $60^\circ$.

\begin{figure}[h!]
	\centering
	\resizebox{\hsize}{!}{\includegraphics{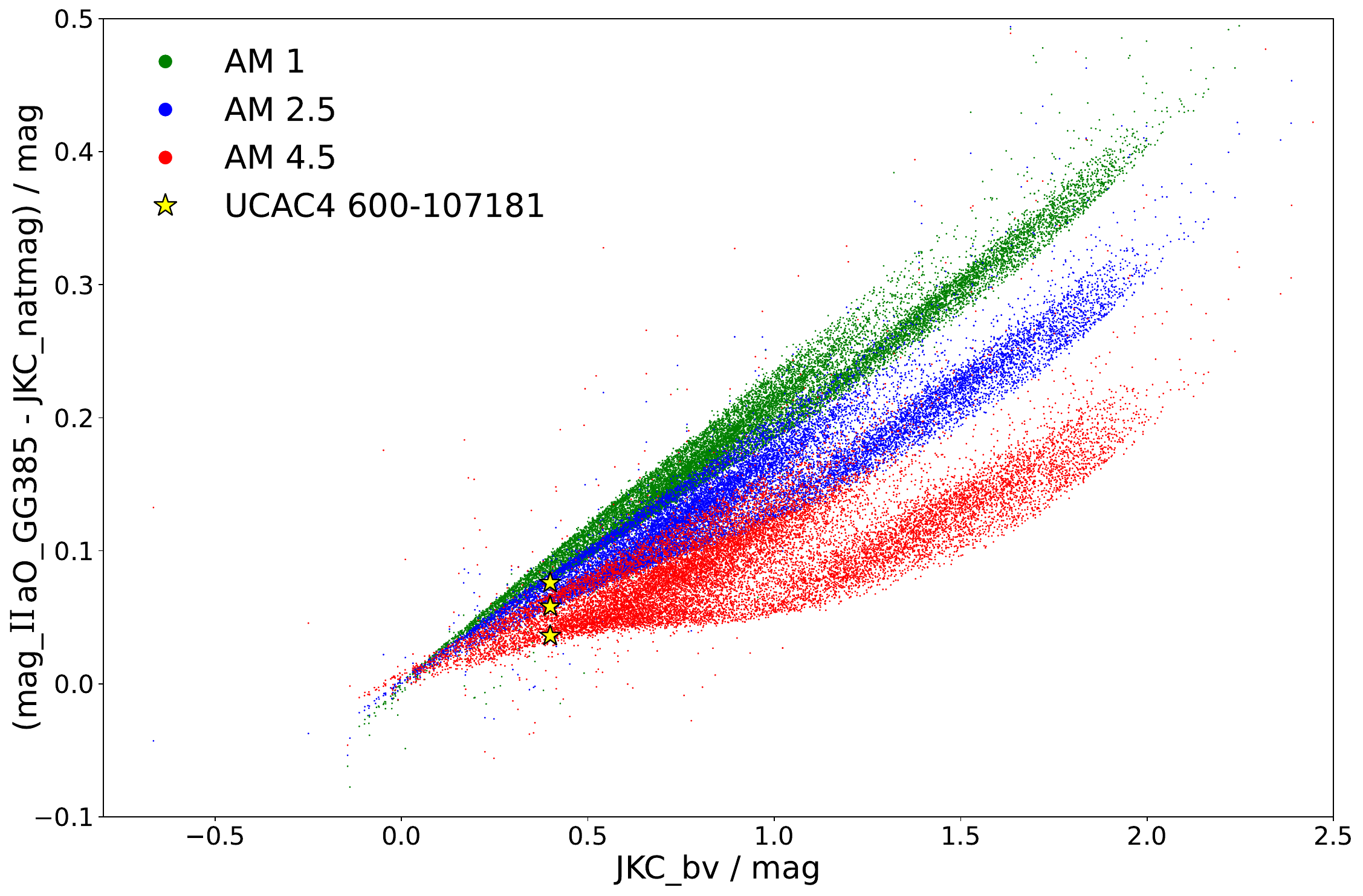}}

	\caption{Dependence on air masses. Difference $\Delta = \mathrm{m}_\mathrm{synth} - ( V_\mathrm{JKC} + c (B_\mathrm{JKC} - V_\mathrm{JKC}))$ versus color index for all in APPLAUSE DR3 identified stars in the second scan of plate 12368 (see Fig.\ \ref{fig5}).  $\mathrm{m}_\mathrm{synth}$ is determined for three different air masses.}
	\label{fig7}
\end{figure}

Since the geometric dimensions of photographic plates are larger than most of the CCDs nowadays in use (except for cameras designed for special purposes), noticeable differences in air mass exist for stars at the lower and upper edges of the plates. The stars of the center of plate 12368, which is analyzed in the left panel of Fig. \ref{fig4}, have an air mass of 1.34, and the air mass at the edges ranges from 1.43 to 1.26. The air mass at the center of plate 11626 is 1.33, and at the lower and upper edges, 1.42 and 1.25, respectively. The differences are higher for observations at lower altitudes.

Below the oxygen absorption bands, the main contribution to $\kappa({\lambda})$ is the Raleigh scattering, $\propto 1/\lambda^4$,  which is most prominent in the ultraviolet and the blue part of the spectrum, leading to reddening of the light passing through the atmosphere. We account for this reddening by multiplying the emulsion sensitivity and filter transmission with an air mass-dependent Raleigh scattering for air masses in the range of 1 to 5, as shown in Fig. \ref{figC1}. To this end, we used solar spectral irradiances which are modeled by the SMARTS (Simple Model of the Atmospheric Radiative Transfer of Sunshine), version 2.9.5 \citep{GUEYMARD2001, GUEYMARD2003}\footnote{\url{https://www.nrel.gov/grid/solar-resource/smarts}}. 
Thus, we get a series of slightly shifted spectral sensitivities for each emulsion-filter combination, which we use with GaiaXPy to calculate the synthetic magnitudes depending on the air mass. We detected a change of 0.01 mag for the pv plate and 0.04 mag for the pg plate, based on our random star with a color index of 0.39 mag. However, for a star with higher $B-V$ values, the magnitude of this effect increased to 0.2 mag for the pg plate (Fig.\ \ref{fig7}).

The second effect is lowering the observed intensity, which is called extinction. 
We can not use the synthetic magnitudes calculated with GaiaXPy from the SEDs and the spectral sensitivities to take into account the extinction, because GaiaXPy uses a zero point (of Vega in our case) for air mass zero.

The extinction coefficient of the atmosphere for a certain passband can be determined from a series of observations of the same object in dependence on the altitude, i.e., of the air mass. The extinction coefficient can vary by a factor of two with the specific parameters of the atmosphere. For accurate results, the extinction coefficient should be monitored in parallel with each measurement. However, this was not possible for most of the measurements using photographic plates. Thus, averaged numbers have to be used for the extinction correction\footnote{\url{https://dasch.cfa.harvard.edu/data-guide/}}. This correction has to be applied in the calibration step and not when determining the natural magnitudes of calibration stars in the plates' color systems.

Another effect associated with the air mass is due to the low quantum efficiency of photographic plates, resulting in longer exposure times compared to modern CCDs. Thus, the air mass changes during exposure, which averages extinction and reddening, limiting the accuracy of such measurements. For the plates used in Fig. \ref{fig4}, we find rather small changes in the air mass of the center of the plates of 0.018 (plate 12368, exposure time 4 min) and 0.023 (plate 11626, exposure time 8 min). However, checking a more extreme plate used at a central altitude at the mid-point of the exposure of 15.7 degrees with an exposure time of 2 hours, we get a difference of air masses of 1.

\subsection{Transmission of the optics}
\label{sec:newapps:optics}

The widths of spectral sensitivities of emulsions-filter combinations used in photoplate photometry are pretty much comparable to the widths of the Johnson-Cousins filters nowadays used with modern detectors. Therefore, the spectral transmission of the optics affects both to the same extent.

In simple applications presented in many textbooks of observational astronomy, it is assumed that the optics have no fluctuations in their transmission over the wavelength range of the passband filters. This obviously is a desirable goal for the design of telescope optics. However, if the optics are not perfect in that sense, their response functions applying standard Johnson-Cousins (or other) filters might not perfectly match the Johnson-Cousins passbands, which needs to be corrected, see for instance \cite{stetson2013}.

The optics used for photoplate observations are astrographs, Schmidt telescopes, and, for smaller apertures, a triplet or quadruplet lens system. In our comparison of natural magnitudes presented in Section \ref{sec:comparison:color_index}, we used plates exposed in a Schmidt mirror optics. Assuming that the effects of reflecting Aluminum surfaces and Schmidt correction plates in the spectral range of the two emulsions are smaller than the typical statistical errors, we did not include the spectral transmission of the optics in this investigation of the method presented in this paper. To get an estimate of the error of our analysis, we compared plates with the emulsion-filter combinations used in Section \ref{sec:comparison:color_index} being exposed in different optical systems. The APLLAUSE archive contains data from plates with these emulsion-filter combinations exposed in different Schmidt telescopes, which we used to check the influence of the optics. The revealed effects are in the range of 0.1 mag.

In the final application of our method, we need to add the spectral characteristics of the optics in the same way as the transmission of the atmosphere.

\subsection{Stars in different luminosity classes}
\label{sec:newapps:luminosity_classes}

\begin{figure}[h!]
	\centering
	\resizebox{\hsize}{!}{\includegraphics{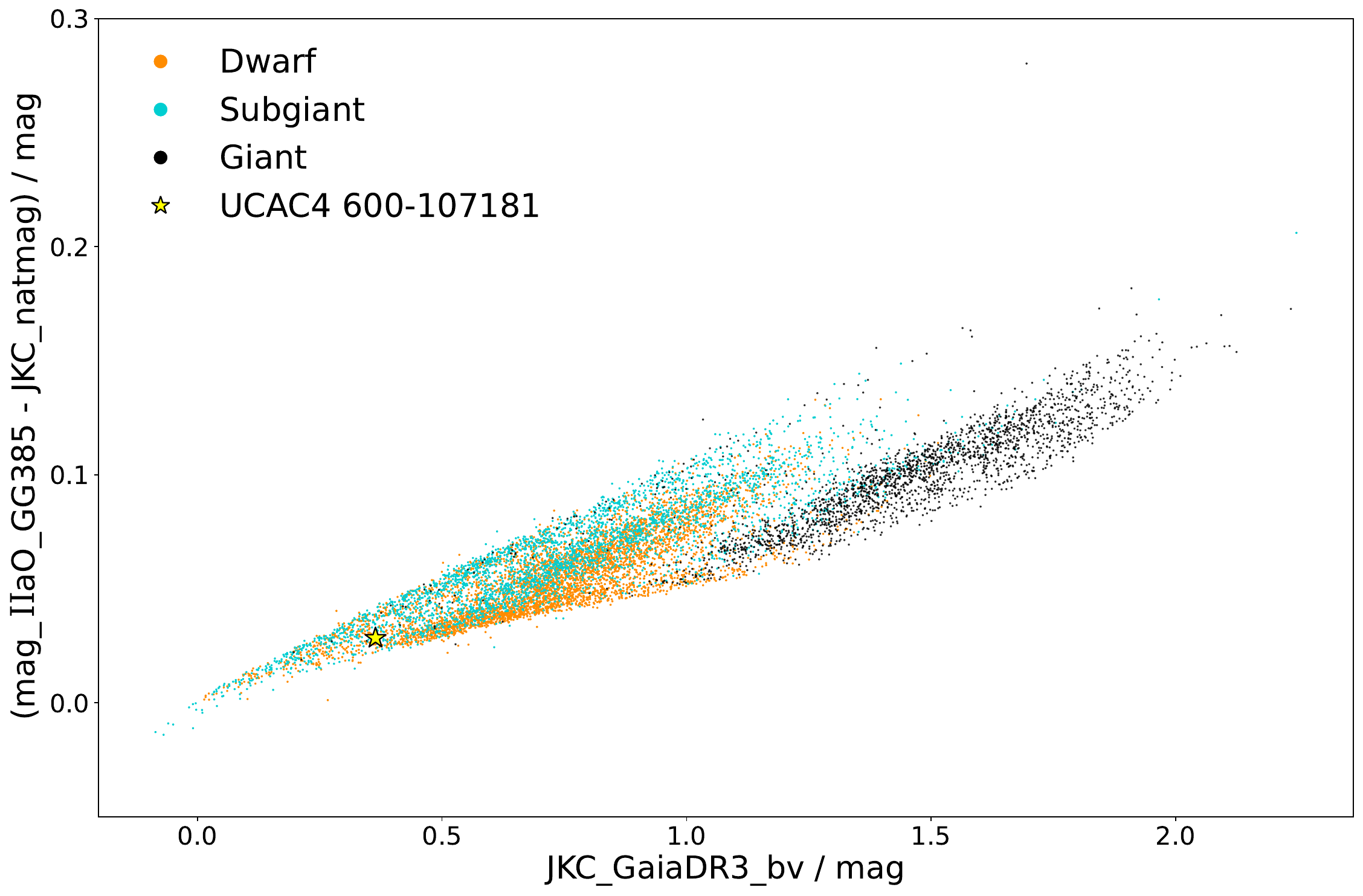}}
	
	\caption{Difference $\Delta = \mathrm{m}_\mathrm{synth} - m_\mathrm{JKC}$ versus color index $\mathrm{JKC\_bv} = B_\mathrm{JKC} - V_\mathrm{JKC\_V} $ for all in APPLAUSE DR3 identified, cleaned and with Gaia DR3 matched stars using synthetic Johnson-Cousins magnitudes from Gaia DR3 for
	plate 12368, Großer Schmidt-Spiegel, Hamburg Observatory, 12.12.1968, Kodak IIa-O with filter GG13(=385), scan 14850 (color term 0.826). Dwarfs and giants, classified using $\log g$ from Gaia DR3, are shown in different colors.}

	\label{fig8}
\end{figure}

In Fig.\ \ref{fig8}, the differences $\Delta = \mathrm{m}_\mathrm{synth} - m_\mathrm{JKC}$ versus color index of one scan of the blue plate 12368 are shown (see also Fig. \ref{fig6}) in more detail, revealing a fine structure.

In the color term approach, the natural magnitude of all stars with an identical color index will be determined with an identical correction. However, the SEDs of dwarfs and giants of identical color index might differ. We classified the stars shown in Fig.\ \ref{fig8} as dwarfs with $\log g \ge 4$, sub-giants with $4 > \log g \ge 3$, and giants with $\log g < 3$ \citep{rodrigo2024} and marked them in different colors. The fine structures in Fig.\ \ref{fig6} and \ref{fig8} are due to stars in different luminosity classes! 

The interesting result is that our method can clearly discriminate dwarfs and giants of identical color index. The difference in the natural magnitudes of the dwarfs and giants in plate 12368 is in a range of 0.1 mag. We should point out that we were not able to separate stars of different luminosity classes in plots of the APPLAUSE data (see Fig.\ \ref{fig5}) because of their erroneous magnitudes and color indices!

\section{Discussion and outlook}
\label{sec:discussion}

The availability of spectral energy distributions of millions of stars in the latest \textit{Gaia} data release opens the chance for an improvement of the photometric calibration of photographic plates by calculating synthetic magnitudes of calibration stars in the individual photometric system of each plate.

We demonstrated proof of concept, mainly analyzing plates with Kodak emulsion, as they have been the ones with easy access to spectral sensitivities. Plates with Agfa Astro Z emulsion, the most frequent in the dataset presented in Fig.\ \ref{fig2}, have been considered for analysis as well, but due to conflicting reports on its spectral sensitivities over the years, it will be addressed in future studies.
The synthetic magnitudes method allows atmospheric reddening and spectral characteristics of the optics to be accounted for. Furthermore, the SEDs of stars belonging to different luminosity classes but with the same colors differ. The presented approach can account for this, whereas the color term method cannot.

Our analysis reveals systematic inconsistencies of the color term method using a linear conversion of the color index of stars from a reference catalog. We clearly identified non-linear errors, too, as well as problems in determining the best color term by the methods used in APPLAUSE. 

Conversion of star colors between different color systems using a Taylor expansion of color indices is a well-established procedure in CCD broadband photometry, see for example \cite{jordi2006}. However, such conversions are always calculated for color systems, which cover nearly identical wavelength ranges with several -- more than one -- passbands each. I.e., a color index in one system -- a ''color'' -- is transformed into a color index of the other system.

A photographic plate, however, is not a complete color system; it just covers one passband. One could consider different emulsions from Kodak, Agfa, Orwo, or others as color systems of a certain series of photographic plates. But to our knowledge, no attempt has been made so far to transform the color system of several emulsions, covering an entire wide spectral range, into a common CCD color system. I.e., all former efforts (APPLAUSE, DASCH, NACD, \ldots ) convert a color index of the reference system to a single passband of each photographic plate. Due to the lack of at least a second passband at a different wavelength, it is obvious to expect systematic errors in dependence on the color index of stars. This aligns precisely with our findings, as illustrated in Fig. \ref{fig5}. One of the major problems of such transformation into the color systems of photographic plates seems to be the determination of the correct color terms!

Laycock mentioned a possible ''slight variation of the color term with the air mass across a wide field plate'' \citep{Laycock2010}, but the announced paper, where they wanted to study this effect, was never published. With the improved photometry of DASCH, they determined an ''effective color response'' (i.e., effective color term) within each of up to 9 annular bins \citep{Tang2013}. However, the air mass does not vary in annular bins across the plate!  A similar idea was used in a preliminary analysis of a few plates from the Sonneberg archive by \cite{Vogt2004}. In each of a series of 60 plates, 11 sub-fields of approximately $2.8^\circ \times 2.8^\circ$ were analyzed by separately fitting color terms. In both studies, the color term varies across a single plate in a range of approximately 0.2, and no systematic correlation of the color terms with geometry, neither annular bins nor distance to the center of the plate nor upper or lower edge, was found! Both studies thus undermine the concept of color terms as a spectral property of an emulsion and, instead, use it as an additional free fitting parameter without any physical meaning. A change of the color term in the range of 0.2 leads to a magnitude change of 0.4 mag for a calibration star with a color index of 2 mag, which is not acceptable for calibration stars!

It is our idea to implement our method of determining the natural magnitude of calibration stars into the open-source code PyPlate and validate it with plates from the DASCH, APPLAUSE, and Sonneberg plate archives. A crucial point of our method is the availability of spectral sensitivities of the photographic plates and transmissions of the filters and optics used in the 20th century. We are aware that we might not be able to find these data for all the emulsions, filters, and optics, but we hope to obtain at least approximate sensitivity curves.

\begin{acknowledgements}
    In this study, we made extensive use of the APPLAUSE archive. It is worth mentioning that the available data at this archive contains all the information about processing the plate, e.g., color terms, parameters used for processing the data, flags for quality checks, a list of non-calibrated sources, etc. We were not able to get this information from the DASCH archive or the NADC archive. APPLAUSE is outstanding in data provenance!

    This work has made use of the Python package GaiaXPy, developed and maintained by members of the \textit{Gaia} Data Processing and Analysis Consortium (DPAC) and, in particular, Coordination Unit 5 (CU5), and the Data Processing Center located at the Institute of Astronomy, Cambridge, UK (DPCI).
    Additionally, this research has used the SVO Filter Profile Service "Carlos Rodrigo," funded by MCIN/AEI/10.13039/501100011033/ through grant PID2020-112949GB-I00.

    This work is funded by the Deutsche Forschungsgemeinschaft (DFG, German Research Foundation) – Projektnummer 543340404.
\end{acknowledgements}

\bibliographystyle{aa} 
\bibliography{raouph_ref}

@INPROCEEDINGS{Schrimpf2024,
	author       = {{Schrimpf}, Andreas},
	title       = {History of the Observation of Stars},
	booktitle	= {Encyclopedia of Astrophysics 1st edition},
	year         = 2025,
	publisher    = {Elsevier },
	editor		= {I. Mandel},
	volume		= {2},
	pages		= {1-12},
	doi =		 {10.1016/B978-0-443-21439-4.00041-9}
}

@ARTICLE{Hudec2019,
	author = {{Hudec}, Rene},
	title = "{Astronomical photographic data archives: Recent status}",
	journal = {Astron. Nachr.},
	year = 2019,
	month = aug,
	volume = {340},
	number = {7},
	pages = {690-697},
	doi = {10.1002/asna.201913676}
}

@ARTICLE{Shang2024,
       author = {{Shang}, Zheng-Jun and {Yu}, Yong and {Wang}, Liang-Liang and {Yang}, Mei-Ting and {Yang}, Jing and {Shen}, Shi-Yin and {Liu}, Min and {Xu}, Quan-Feng and {Cui}, Chen-Zhou and {Fan}, Dong-Wei and {Tang}, Zheng-Hong and {Zhao}, Jian-Hai},
        title = "{Digitization of Astronomical Photographic Plates of China and Astrometric Measurement of Single-exposure Plates}",
      journal = {Res. Astron. Astrophys},
         year = 2024,
        month = may,
       volume = {24},
       number = {5},
          eid = {055010},
        pages = {055010},
          doi = {10.1088/1674-4527/ad339d},
}

@ARTICLE{Robert2021,
       author = {{Robert}, V. and {Desmars}, J. and {Lainey}, V. and {Arlot}, J. -E. and {Perlbarg}, A. -C. and {Horville}, D. and {Aboudarham}, J. and {Etienne}, C. and {Gu{\'e}rard}, J. and {Ilovaisky}, S. and {Khovritchev}, M.~Y. and {Le Poncin-Lafitte}, C. and {Le Van Suu}, A. and {Neiner}, C. and {Pascu}, D. and {Poirier}, L. and {Schneider}, J. and {Tanga}, P. and {Valls-Gabaud}, D.},
        title = "{The NAROO digitization center. Overview and scientific program}",
      journal = {\aap},
         year = 2021,
        month = aug,
       volume = {652},
          eid = {A3},
        pages = {A3},
          doi = {10.1051/0004-6361/202140472},
}

@unpublished{Yuan2024,
        author = {Yuan, Haibo},
        title = "{Astrometry and photometry of digitized images of Chinese plates}",
        date   = {2024-06-18/2024-06-21},
        howpublished = {Conference session},
        addendum     = {PDPP Workshop 2024: Technical Requirements for Digitizing Direct Photographic Plates, Shanghai, China},
        year={2024},
        note = "{PDPP Workshop 2024: Technical Requirements for Digitizing Direct Photographic Plates, Shanghai, China}"
}

@INPROCEEDINGS{Grindlay2009,
       author = {{Grindlay}, J. and {Tang}, S. and {Simcoe}, R. and {Laycock}, S. and {Los}, E. and {Mink}, D. and {Doane}, A. and {Champine}, G.},
        title = "{DASCH to Measure (and preserve) the Harvard Plates: Opening the {\ensuremath{\sim}}100-year Time Domain Astronomy Window}",
    booktitle = {Preserving Astronomy's Photographic Legacy: Current State and the Future of North American Astronomical Plates},
         year = 2009,
       editor = {{Osborn}, W. and {Robbins}, L.},
       series = {ASP Conference Series},
       volume = {410},
        month = aug,
        pages = {101},
}

@ARTICLE{Laycock2010,
       author = {{Laycock}, S. and {Tang}, S. and {Grindlay}, J. and {Los}, E. and {Simcoe}, R. and {Mink}, D.},
        title = "{Digital Access to a Sky Century at Harvard: Initial Photometry and Astrometry}",
      journal = {\aj},
         year = 2010,
        month = oct,
       volume = {140},
       number = {4},
        pages = {1062-1077},
          doi = {10.1088/0004-6256/140/4/1062},
}

@article{Enke2024,
   		author={Enke, Harry and Tuvikene, Taavi and Groote, Detlef and Edelmann, Heinz and Heber, Ulrich},
   		title={Archives of Photographic PLates for Astronomical USE (APPLAUSE): Digitisation of astronomical plates and their integration into the International Virtual Observatory},
   		volume={687},
   		url={http://dx.doi.org/10.1051/0004-6361/202348793},
   		DOI={10.1051/0004-6361/202348793},
   		journal={\aap},
   		year={2024},
  		month=jul, pages={A165} 
}

@INPROCEEDINGS{Kroll2009,
       author = {{Kroll}, Peter},
        title = "{Real and Virtual Heritage - The Plate Archive of Sonneberg Observatory - Digitisation, Preservation and Scientific Programme}",
    booktitle = {Cultural Heritage of Astronomical Observatories: From Classical Astronomy to Modern Astrophysics},
         year = 2009,
       editor = {{Wolfschmidt}, Gudrun},
        month = jan,
        pages = {311-315},
}

@INPROCEEDINGS{Tuvikene2014,
      	author = {{Tuvikene}, T. and {Edelmann}, H. and {Groote}, D. and {Enke}, H.},
        title = "{Work flow for plate digitization, data extraction and publication}",
   		booktitle = {Astroplate 2014},
        year = 2014,
        month = jan,
        pages = {127},
       	adsurl = {https://ui.adsabs.harvard.edu/abs/2014aspl.conf..127T}
}

@inproceeding{Tuvikene2019,
  		author={{Tuvikene}, Taavi and {APPLAUSE Collaboration} },
 		title= "{PyPlate: a software package for processing digitized astronomical photographic plates}",
 		booktitle    = {{L}arge {S}urveys with {S}mall {T}elescopes: Past, Present and Future (Astroplate III)},
 		note         = {{B}amberg, Germany, March 11–13, 2019},
 		year         = {2019},
  		pages={11--13},
  		year={2019}
}

@inproceeding{Kroll2019,
	author       = {Kroll, Peter},
	title        = {Photographic and Digital Sky Surveys at Sonneberg Observatory},
	booktitle    = {{L}arge {S}urveys with {S}mall {T}elescopes: Past, Present and Future (Astroplate III)},
	note         = {{B}amberg, Germany, March 11–13, 2019},
	year         = {2019},
	url          = {https://www.plate-archive.org/cms/wp-content/uploads/2019/04/Kroll_PhotographicAndDigitalSkySurveysAtSonnebergObservatory.pdf},
	urldate      = {2025-07-17}
}

@INPROCEEDINGS{Osborn2024,
	author = {{Osborn}, Wayne and {Bauer}, Amanda},
	title = "{Yerkes treasures: The largest photographic plate ever used in astronomy}",
	booktitle = {American Astronomical Society Meeting Abstracts},
	year = 2024,
	series = {American Astronomical Society Meeting Abstracts},
	volume = {244},
	month = jun,
	eid = {122.01},
	pages = {122.01},
	adsurl = {https://ui.adsabs.harvard.edu/abs/2024AAS...24412201O}
}

@BOOK{Jones1971,
	author = {{Jones}, Bessie Zaban and {Boyd}, Lyle Gifford},
	title = "{The Harvard College Observatory: the first four directorships, 1839-1919}",
	year = 1971,
	doi = {10.4159/harvard.9780674418806},
	adsurl = {https://ui.adsabs.harvard.edu/abs/1971hcof.book.....J}
}

@ARTICLE{Braeuer1999,
	author = {{Br{\"a}uer}, Hans-J{\"u}rgen and {H{\"a}usele}, Inge and {L{\"o}chel}, Klaus and {Polko}, Norbert},
	title = "{Structure and content of the Sonneberg Plate Archive}",
	journal = {Acta Historica Astronomiae},
	year = 1999,
	month = jan,
	volume = {6},
	pages = {70-73},
	adsurl = {https://ui.adsabs.harvard.edu/abs/1999AcHA....6...70B}
}

@INPROCEEDINGS{Spasovic2016,
	author = {{Spasovic}, Milan and {Dersch}, Christian and {Lange}, Christian and {Jovanovic}, Dragan and {Schrimpf}, Andreas},
	title = "{Sonneberg Sky Patrol Archive - Photometric Analysis}",
   	booktitle = {Proceedings Astroplate 2016},
	year = 2016,
	pages = {67},
	url = {https://www.astroplate.cz/wp-content/uploads/Proceedings/AstroplateProceedings2016.pdf},
}

@ARTICLE{Martin1976,
	author = {{Martin}, W.~L.},
	title = "{Photographic Photometry using the twin Refractor at SAAO (Sutherland)}",
	journal = {Monthly Notes of the Astronomical Society of South Africa},
	year = 1976,
	month = jan,
	volume = {35},
	pages = {97},
	adsurl = {https://ui.adsabs.harvard.edu/abs/1976MNSSA..35...97M}
}

@ARTICLE{Anderson2004,
       author = {{Anderson}, S.~R. and {Engels}, D.},
        title = "{A short history of Hamburg Observatory}",
      journal = {Journal of the British Astronomical Association},
         year = 2004,
        month = apr,
       volume = {114},
        pages = {78-87},
}

@article{Vallenari2023,
      author        = {{Gaia Collaboration (Vallenari, A. et al.)}},
      xauthor =     {Vallenari, A. and Brown, A.G.A. and Prusti, T. and de Bruijne, J.H.J. and Arenou, F. and Babusiaux, C. and Biermann, M. and Creevey, O.L. and Ducourant, C. and Evans, D.W. and Eyer, L. and Guerra, R. and Hutton, A. and Jordi, C. and Klioner, S.A. and Lammers, U.L. and Lindegren, L. and Luri, X. and Mignard, F. and Panem, C. },
      collaboration = {Gaia Collaboration},
      title         = "{Gaia Data Release 3 - Summary of the content and survey
                       properties}",
      journal       = {\aap},
      volume        = {674},
      pages         = {A1},
      year          = {2023},
      doi           = {10.1051/0004-6361/202243940},
}

@article{Montegriffo2023a,
		title = "{Gaia Data Release 3. The Galaxy in your preferred colours. Synthetic photometry from Gaia low-resolution spectra}",
		author = {{Gaia Collaboration (Montegriffo, P. et al.)}},
        xauthor = {{Montegriffo}, P. and {Bellazzini}, M. and {De Angeli}, F. and {Andrae}, R. and {Barstow}, M.~A. and {Bossini}, D. and {Bragaglia}, A. and {Burgess}, P.~W. and {Cacciari}, C. and {Carrasco}, J.~M. and {Chornay}, N. and {Delchambre}, L. and {Evans}, D.~W. and {Fouesneau}, M. and {Fr{\'e}mat}, Y. and {Garabato}, D. and {Jordi}, C. and {Manteiga}, M. and {Massari}, D. and {Palaversa}, L. and {Pancino}, E. and {Riello}, M. and {Ruz Mieres}, D. and {Sanna}, N. and {Santove{\~n}a}, R. and {Sordo}, R. and {Vallenari}, A. and {Walton}, N.~A. and {Brown}, A.~G.~A. and {Prusti}, T. and {de Bruijne}, J.~H.~J. and {Arenou}, F. and {Babusiaux}, C. and {Biermann}, M. and {Creevey}, O.~L. and {Ducourant}, C. and {Eyer}, L. and {Guerra}, R. and {Hutton}, A. and {Klioner}, S.~A. and {Lammers}, U.~L. and {Lindegren}, L. and {Luri}, X. and {Mignard}, F. and {Panem}, C. and {Pourbaix}, D. and {Randich}, S. and {Sartoretti}, P. and {Soubiran}, C. and {Tanga}, P. and {Bailer-Jones}, C.~A.~L. and {Bastian}, U. and {Drimmel}, R. and {Jansen}, F. and {Katz}, D. and {Lattanzi}, M.~G. and {van Leeuwen}, F. and {Bakker}, J. and {Casta{\~n}eda}, J. and {Fabricius}, C. and {Galluccio}, L. and {Guerrier}, A. and {Heiter}, U. and {Masana}, E. and {Messineo}, R. and {Mowlavi}, N. and {Nicolas}, C. and {Nienartowicz}, K. and {Pailler}, F. and {Panuzzo}, P. and {Riclet},},
      	collaboraation = {Gaia Collaboration},
         year = 2023,
        month = jun,
       volume = {674},
          eid = {A33},
        pages = {A33},
          doi = {10.1051/0004-6361/202243709},
		journal = {\aap},
}

@ARTICLE{DeAngeli2023,
       title = "{Gaia Data Release 3. Processing and validation of BP/RP low-resolution spectral data}",
           author = {{Gaia Collaboration (De Angeli, F. et al.)}},
       xauthor = {{De Angeli}, F. and {Weiler}, M. and {Montegriffo}, P. and {Evans}, D.~W. and {Riello}, M. and {Andrae}, R. and {Carrasco}, J.~M. and {Busso}, G. and {Burgess}, P.~W. and {Cacciari}, C. and {Davidson}, M. and {Harrison}, D.~L. and {Hodgkin}, S.~T. and {Jordi}, C. and {Osborne}, P.~J. and {Pancino}, E. and {Altavilla}, G. and {Barstow}, M.~A. and {Bailer-Jones}, C.~A.~L. and {Bellazzini}, M. and {Brown}, A.~G.~A. and {Castellani}, M. and {Cowell}, S. and {Delchambre}, L. and {De Luise}, F. and {Diener}, C. and {Fabricius}, C. and {Fouesneau}, M. and {Fr{\'e}mat}, Y. and {Gilmore}, G. and {Giuffrida}, G. and {Hambly}, N.~C. and {Hidalgo}, S. and {Holland}, G. and {Kostrzewa-Rutkowska}, Z. and {van Leeuwen}, F. and {Lobel}, A. and {Marinoni}, S. and {Miller}, N. and {Pagani}, C. and {Palaversa}, L. and {Piersimoni}, A.~M. and {Pulone}, L. and {Ragaini}, S. and {Rainer}, M. and {Richards}, P.~J. and {Rixon}, G.~T. and {Ruz-Mieres}, D. and {Sanna}, N. and {Sarro}, L.~M. and {Rowell}, N. and {Sordo}, R. and {Walton}, N.~A. and {Yoldas}, A.},
        title = "{Gaia Data Release 3. Processing and validation of BP/RP low-resolution spectral data}",
      journal = {\aap},
         year = 2023,
        month = jun,
       volume = {674},
          eid = {A2},
        pages = {A2},
          doi = {10.1051/0004-6361/202243680},
archivePrefix = {arXiv},
       eprint = {2206.06143},
 primaryClass = {astro-ph.IM},
       adsurl = {https://ui.adsabs.harvard.edu/abs/2023A&A...674A...2D}
}

@article{Bessell2012,
		author = {Michael Bessell and Simon Murphy},
		title = {Spectrophotometric Libraries, Revised Photonic Passbands, and Zero Points for UBVRI, Hipparcos, and Tycho Photometry},		
		doi = {10.1086/664083},
		url = {https://dx.doi.org/10.1086/664083},
		year = {2012},
		month = {feb},
		publisher = {University of Chicago Press},
		volume = {124},
		number = {912},
		pages = {140},
		journal = {PASP},
}

@INPROCEEDINGS{svo2020,
       author = {{Rodrigo}, C. and {Solano}, E.},
        title = "{The SVO Filter Profile Service}",
    booktitle = {XIV.0 Scientific Meeting (virtual) of the Spanish Astronomical Society},
         year = 2020,
        month = jul,
          eid = {182},
        pages = {182},
       adsurl = {https://ui.adsabs.harvard.edu/abs/2020sea..confE.182R}
}

@ARTICLE{Bell1972,
       author = {{Bell}, R.~A.},
        title = "{The application of synthetic spectra to colour system transforma-tions-II. The photographic and photoelectric UBV systems}",
      journal = {\mnras},
         year = 1972,
        month = jan,
       volume = {159},
        pages = {357},
          doi = {10.1093/mnras/159.4.357},
       adsurl = {https://ui.adsabs.harvard.edu/abs/1972MNRAS.159..357B}
}

@ARTICLE{Vogt2004,
       author = {{Vogt}, N. and {Kroll}, P. and {Splittgerber}, E.},
        title = "{A photometric pilot study on Sonneberg archival patrol plates. How many ``constant'' stars are in fact long-term variables?}",
      journal = {\aap},
         year = 2004,
        month = dec,
       volume = {428},
        pages = {925-934},
          doi = {10.1051/0004-6361:20040457},
       adsurl = {https://ui.adsabs.harvard.edu/abs/2004A&A...428..925V}
}

@ARTICLE{Tang2013,
       author = {{Tang}, Sumin and {Grindlay}, Jonathan and {Los}, Edward and {Servillat}, Mathieu},
        title = "{Improved Photometry for the DASCH Pipeline}",
      journal = {\pasp},
         year = 2013,
        month = jul,
       volume = {125},
       number = {929},
        pages = {857},
          doi = {10.1086/671760},
archivePrefix = {arXiv},
       eprint = {1304.7504},
 primaryClass = {astro-ph.IM},
       adsurl = {https://ui.adsabs.harvard.edu/abs/2013PASP..125..857T}
}

@article{GUEYMARD2003,
title = {Direct solar transmittance and irradiance predictions with broadband models. Part I: detailed theoretical performance assessment},
journal = {Solar Energy},
volume = {74},
number = {5},
pages = {355-379},
year = {2003},
issn = {0038-092X},
doi = {https://doi.org/10.1016/S0038-092X(03)00195-6},
url = {https://www.sciencedirect.com/science/article/pii/S0038092X03001956},
author = {Christian A. Gueymard}}

@article{GUEYMARD2001,
title = {Parameterized transmittance model for direct beam and circumsolar spectral irradiance},
journal = {Solar Energy},
volume = {71},
number = {5},
pages = {325-346},
year = {2001},
issn = {0038-092X},
doi = {https://doi.org/10.1016/S0038-092X(01)00054-8},
url = {https://www.sciencedirect.com/science/article/pii/S0038092X01000548},
author = {Christian A. Gueymard}}

@ARTICLE{Agerer1997,
       author = {{Agerer}, Franz and {Huebscher}, Joachim},
        title = "{Photoelectric Minima of Selected Eclipsing Binaries and Maxima of Pulsating Stars}",
      journal = {Information Bulletin on Variable Stars},
         year = 1997,
        month = apr,
       volume = {4472},
        pages = {1},
       adsurl = {https://ui.adsabs.harvard.edu/abs/1997IBVS.4472....1A}
}

@ARTICLE{Anki2021,
       author = {{Aoki}, Tsutomu and {Soyano}, Takao and {Nakajima}, Koichi and {Miyauchi}, Nagako and {Mori}, Yuki and {Tarusawa}, Ken'ichi and {Kobayashi}, Naoto and {Furusawa}, Junko and {Ichikawa}, Shin-Ichi and {Smoka Group}},
        title = "{Digitization of Photographic Plates at Kiso Observatory}",
      journal = {Astronomical Herald},
         year = 2021,
        month = jul,
       volume = {8},
        pages = {523-533},
       adsurl = {https://ui.adsabs.harvard.edu/abs/2021AstHe.114..523A}
}

@ARTICLE{Moro2000,
       author = {{Moro}, D. and {Munari}, U.},
        title = "{The Asiago Database on Photometric Systems (ADPS). I. Census parameters for 167 photometric systems}",
      journal = {\aaps},
         year = 2000,
        month = dec,
       volume = {147},
        pages = {361-628},
          doi = {10.1051/aas:2000370},
       adsurl = {https://ui.adsabs.harvard.edu/abs/2000A&AS..147..361M}
}

@misc{noirlab_schott_filters,
  author       = {{NSF NOIRLab}},
  title        = {Schott Glass Filters – Hydra/CTIO},
  howpublished = {\url{https://noirlab.edu/science/programs/ctio/filters/Hydra/Schott-Glass-Filters}},
  year         = 2021
}

@ARTICLE{Dokuchaeva1983,
       author = {{Dokuchaeva}, O.~D. and {Birulya}, T.~A. and {Toropova}, M.~S.},
        title = "{The Study of Hypersensitizing of Photographic Materials in the Photolaboratory of the Sternberg State Astronomical Institute}",
      journal = {Nauchnye Informatsii},
         year = 1983,
        month = jan,
       volume = {54},
        pages = {160},
       adsurl = {https://ui.adsabs.harvard.edu/abs/1983NInfo..54..160D}
}

@misc{STScI_GSC_Surveys,
  author       = {{Space Telescope Science Institute}},
  title        = {Guide Star Catalog and Sky Surveys},
  howpublished = {\url{https://gsss.stsci.edu/SkySurveys/Surveys.htm}},
  year         = {2016}
}

@book{Kodaknotebook1967,
  author       = {{Eastman Kodak Handbook}},
  title        = {Kodak Plates and Films for Science and Industry},
  edition      = {1st},
  series       = {Kodak Publication P-9},
  year         = {1967},
  address      = {Rochester, N.Y.},
  publisher    = {Eastman Kodak Company},
  pages        = {32},
}

@INCOLLECTION{stetson2013,
       author = {{Stetson}, Peter B.},
        title = "{Astronomical Photometry}",
    booktitle = {Planets, Stars and Stellar Systems. Volume 2: Astronomical Techniques, Software and Data},
         year = 2013,
       editor = {{Oswalt}, Terry D. and {Bond}, Howard E.},
        pages = {1},
          doi = {10.1007/978-94-007-5618-2_1},
       adsurl = {https://ui.adsabs.harvard.edu/abs/2013pss2.book....1S}
}

@ARTICLE{Zacharias2013,
       author = {{Zacharias}, N. and {Finch}, C.~T. and {Girard}, T.~M. and {Henden}, A. and {Bartlett}, J.~L. and {Monet}, D.~G. and {Zacharias}, M.~I.},
        title = "{The Fourth US Naval Observatory CCD Astrograph Catalog (UCAC4)}",
      journal = {\aj},
         year = 2013,
        month = feb,
       volume = {145},
       number = {2},
          eid = {44},
        pages = {44},
          doi = {10.1088/0004-6256/145/2/44},
archivePrefix = {arXiv},
       eprint = {1212.6182},
 primaryClass = {astro-ph.IM},
       adsurl = {https://ui.adsabs.harvard.edu/abs/2013AJ....145...44Z}
}

@ARTICLE{Høg2000,
       author = {{H{\o}g}, E. and {Fabricius}, C. and {Makarov}, V.~V. and {Urban}, S. and {Corbin}, T. and {Wycoff}, G. and {Bastian}, U. and {Schwekendiek}, P. and {Wicenec}, A.},
        title = "{The Tycho-2 catalogue of the 2.5 million brightest stars}",
      journal = {\aap},
         year = 2000,
        month = mar,
       volume = {355},
        pages = {L27-L30},
       adsurl = {https://ui.adsabs.harvard.edu/abs/2000A&A...355L..27H}
}

@dataset{Henden2016,
       author = {{Henden}, A.~A. and {Templeton}, M. and {Terrell}, D. and {Smith}, T.~C. and {Levine}, S. and {Welch}, D.},
        title = "{VizieR Online Data Catalog: AAVSO Photometric All Sky Survey (APASS) DR9 (Henden+, 2016)}",
 howpublished = {VizieR On-line Data Catalog: II/336.  Originally published in: 2015AAS...22533616H},
         year = 2016,
        month = jan,
          eid = {II/336},
       adsurl = {https://ui.adsabs.harvard.edu/abs/2016yCat.2336....0H}
}

@ARTICLE{Brown2011,
       author = {{Brown}, Timothy M. and {Latham}, David W. and {Everett}, Mark E. and {Esquerdo}, Gilbert A.},
        title = "{Kepler Input Catalog: Photometric Calibration and Stellar Classification}",
      journal = {\aj},
         year = 2011,
        month = oct,
       volume = {142},
       number = {4},
          eid = {112},
        pages = {112},
          doi = {10.1088/0004-6256/142/4/112},
archivePrefix = {arXiv},
       eprint = {1102.0342},
 primaryClass = {astro-ph.SR},
       adsurl = {https://ui.adsabs.harvard.edu/abs/2011AJ....142..112B}
}

@dataset{Lasker2021,
       author = {{Lasker}, B. and {Lattanzi}, M.~G. and {McLean}, B.~J. and {Bucciarelli}, B. and {Drimmel}, R. and {Garcia}, J. and {Greene}, G. and {Guglielmetti}, F. and {Hanley}, C. and {Hawkins}, G. and {Laidler}, V.~G. and {Loomis}, C. and {Meakes}, M. and {Mignani}, R. and {Morbidelli}, R. and {Morrison}, J. and {Pannunzio}, R. and {Rosenberg}, A. and {Sarasso}, M. and {Smart}, R.~L. and {Spagna}, A. and {Sturch}, C.~R. and {Volpicelli}, A. and {White}, R.~L. and {Wolfe}, D. and {Zacchei}, A.},
        title = "{VizieR Online Data Catalog: The Guide Star Catalog, Version 2.4.2 (GSC2.4.2) (STScI, 2020)}",
 howpublished = {VizieR On-line Data Catalog: I/353.  Originally published in: 2008AJ....136..735L},
         year = 2021,
        month = oct,
          eid = {I/353},
       adsurl = {https://ui.adsabs.harvard.edu/abs/2021yCat.1353....0L}
}

@INCOLLECTION{johnson1963,
	author = {{Johnson}, H. L.},
	title = "{Photometric Systems}",
	booktitle = {Basic Astronomical Data},
	publisher    = {Chicago University Press},
	year = 1963,
	editor = {{Strand}, K.A.},
	pages = {204},
}

@ARTICLE{fukugita1996,
	author = {{Fukugita}, M. and {Ichikawa}, T. and {Gunn}, J.~E. and {Doi}, M. and {Shimasaku}, K. and {Schneider}, D.~P.},
	title = "{The Sloan Digital Sky Survey Photometric System}",
	journal = {\aj},
	year = 1996,
	month = apr,
	volume = {111},
	pages = {1748},
	doi = {10.1086/117915},
	adsurl = {https://ui.adsabs.harvard.edu/abs/1996AJ....111.1748F}
}

@ARTICLE{jordi2006,
	author = {{Jordi}, K. and {Grebel}, E.~K. and {Ammon}, K.},
	title = "{Empirical color transformations between SDSS photometry and other photometric systems}",
	journal = {\aap},
	year = 2006,
	month = dec,
	volume = {460},
	number = {1},
	pages = {339-347},
	doi = {10.1051/0004-6361:20066082},
	archivePrefix = {arXiv},
	eprint = {astro-ph/0609121},
	primaryClass = {astro-ph},
	adsurl = {https://ui.adsabs.harvard.edu/abs/2006A&A...460..339J}
}

@ARTICLE{rodrigo2024,
	author = {{Rodrigo}, Carlos and {Cruz}, Patricia and {Aguilar}, John F. and {Aller}, Alba and {Solano}, Enrique and {G{\'a}lvez-Ortiz}, Maria Cruz and {Jim{\'e}nez-Esteban}, Francisco and {Mas-Buitrago}, Pedro and {Bayo}, Amelia and {Cort{\'e}s-Contreras}, Miriam and {Murillo-Ojeda}, Raquel and {Bonoli}, Silvia and {Cenarro}, Javier and {Dupke}, Renato and {L{\'o}pez-Sanjuan}, Carlos and {Mar{\'\i}n-Franch}, Antonio and {de Oliveira}, Claudia Mendes and {Moles}, Mariano and {Taylor}, Keith and {Varela}, Jes{\'u}s and {Rami{\'o}}, H{\'e}ctor V{\'a}zquez},
	title = "{Photometric segregation of dwarf and giant FGK stars using the SVO Filter Profile Service and photometric tools}",
	journal = {\aap},
	year = 2024,
	month = sep,
	volume = {689},
	eid = {A93},
	pages = {A93},
	doi = {10.1051/0004-6361/202449998},
	archivePrefix = {arXiv},
	eprint = {2406.03310},
	primaryClass = {astro-ph.SR},
	adsurl = {https://ui.adsabs.harvard.edu/abs/2024A&A...689A..93R}
}

\begin{appendix}
\onecolumn

\section{Details about the plates used in Fig.\ \ref{fig2}}
\label{sec:appendix:A}

\begin{table*}[h!]
	\centering
	\caption{Details from APPLAUSE DR3 about all 1\,181 scans of 606 photographic plates containing star UCAC4 600-107181, collected during 1923–1955. The first column shows the emulsions and filter combinations, the second and third columns correspond to the number of plates and the range of color term range, respectively, and the fourth column lists the emulsion type pv (103a-D) or pg (IIa-O).} 
	\begin{tabular}{lcrc}
 \hline \hline		
		Emulsion + Filter & Number & Color Term Range  & Emulsion Type\\
		\hline

		Agfa Astro Z  & 205 & 0.57 \,--\, 1.14 & pg \\
		\textsuperscript{*}Agfa Astro V  & 160 & -0.44 \,--\, 0.91 & pg \\
		Kodak Oa-O  & 79 & 0.64 \,--\, 1.00 & pg \\
		Kodak IIa-O + GG13(=385) & 32 & 0.55 \,--\, 1.16 & pg \\
		Herzog panchromodux + OG5(=550) & 12 & -0.23 \,--\, -0.13  & pv \\
		Kodak 103a-D + GG11  & 22 & -0.09 \,--\, 0.01 & pv \\
		Kodak 103a-O + UG1  & 18 & 0.71 \,--\, 1.85 & pg\\
		Unknown & 13 & 0.50 \,--\, 1.02  & pg \\
		Agfa Astro Z Spiegel & 12 & 0.60 \,--\, 0.71  & pg \\
		Kodak 103a-O  & 12 & 0.44 \,--\, 0.93 & pg \\
		\textsuperscript{*}Perutz Phototech. B + BG3 + GG13 & 10 & 0.07 \,--\, 0.98 & pg \\
		Perutz Astro + GG13(=385) & 7 & 0.70 \,--\, 0.74  & pg  \\
		\textsuperscript{*}Kodak 103a-D + UG1 & 7 & 0.00 \,--\, 1.17 & pg  \\
		Agfa Astro  & 3 & 0.89 \,--\, 0.90 & pg \\
		Agfa Astro ZS  & 3 & 0.90 \,--\, 0.94 & pg \\
		Kodak IIa-O  & 2 & 0.67 \,--\, 0.72 & pg \\
		Perutz Astro  & 2 & 0.99 \,--\, 1.03 & pg \\
		\textsuperscript{*}Kodak 103a-D & 1 & 0.63 & pg! \\
		Eastman OaO & 1 & 0.60 & pg \\
		Agfa isorapid  & 1 & 0.95 & pg \\
		Kodak IIa-O + UG1  & 1 & 0.94 & pg\\
		Agfa Astro O  & 1 & 0.80 & pg \\
		Unknown + OG5(=550) & 1 & -0.21 & pv\\
		Perutz Phototech. B & 1 & 0.88  & pg\\
		\bottomrule
	\end{tabular}
\end{table*}

In the APPLAUSE DR3, some photographic plates were scanned twice, with the second scan rotated by 90° to correct scanner-induced geometric distortions. With the use of high-precision \textit{Gaia} astrometry in APPLAUSE DR4, this procedure was no longer required. The photometric measurements from the two scans differed, so both were included in our analysis. In total, 606 plates were selected for this study, most have a second scan, resulting in 1\,181 scans.

In cases where reliable information on the sensitivity of the emulsion was lacking or not specified, we classified the plates according to the measured color terms: values above 0.5 were considered blue-sensitive (pg) and values below this threshold visible-light-sensitive (pv). 

In Fig. 2, panel 1, four notable mismatches were identified (two points are duplicated due to the second scan), whose categories are marked with an asterisk in the table. 
Two plates labeled blue sensitive, including plate 8147 with Agfa Astro V emulsion and plate 11119 with Kodak 103a-D emulsion behind the UG1 Schott filter, show color terms near zero, although they are expected to show values closer to 1.
Plate 69797 with Perutz Phototech. B emulsion with the BG3 + GG13 filter combination, which restricts transmission to the blue spectral band \citep{Agerer1997}, takes a value of zero once within ten repetitions in the data set. Moreover, plate 4662 with the Kodak 103a-D emulsion, known to be yellow/red sensitive, shows a color term of 0.63, whereas it is expected to be close to zero.

\section{Data cleaning and refinement for visualization in Fig.\ \ref{fig5}}
\label{sec:appendix:B}

The APPLAUSE DR3 lists 214\,821 calibrated sources in the blue plate 12368 and 336\,976 calibrated sources in the visible plate 11626.

In the cleaning process, we skipped all the sources without a UCAC4 id, which we need for matching with \textit{Gaia} sources. The APPLAUSE database lists parameters and flags of the different steps of the analysis. In the next step, we filtered out the sources with ambiguous classification from the source extractor (sourceextractor\_flag not 0) and flagged sources due to uncertain photometric results (phot\_plate\_flags not 0). And, we only kept sources with UCAC4 photometric data in the $B$ and $V$ bands and in the cat\_natmag (magnitude in the natural photometric plate system) column.

In the following step, we matched the UCAC4 sources with stars from the \textit{Gaia} DR3 catalog using \textit{Gaia} positional and proper motion data, as described on the \textit{Gaia} web pages\footnote{\url{https://www.cosmos.esa.int/web/gaia-users/archive/combine-with-other-data}}. All multiple matches of \textit{Gaia} sources to single plate sources (due to the higher spatial resolution of \textit{Gaia}) were removed. Variable stars were excluded as the last stage of data cleaning, and finally we ended up with 15\,346 sources for the blue plate 12368 and 16\,807 sources for the visible plate 11626.

\section {Effect of the air masses}
\label {sec:appendix:C}

\begin{figure*}[h!]
	\centering
	\includegraphics[width=0.495\hsize]{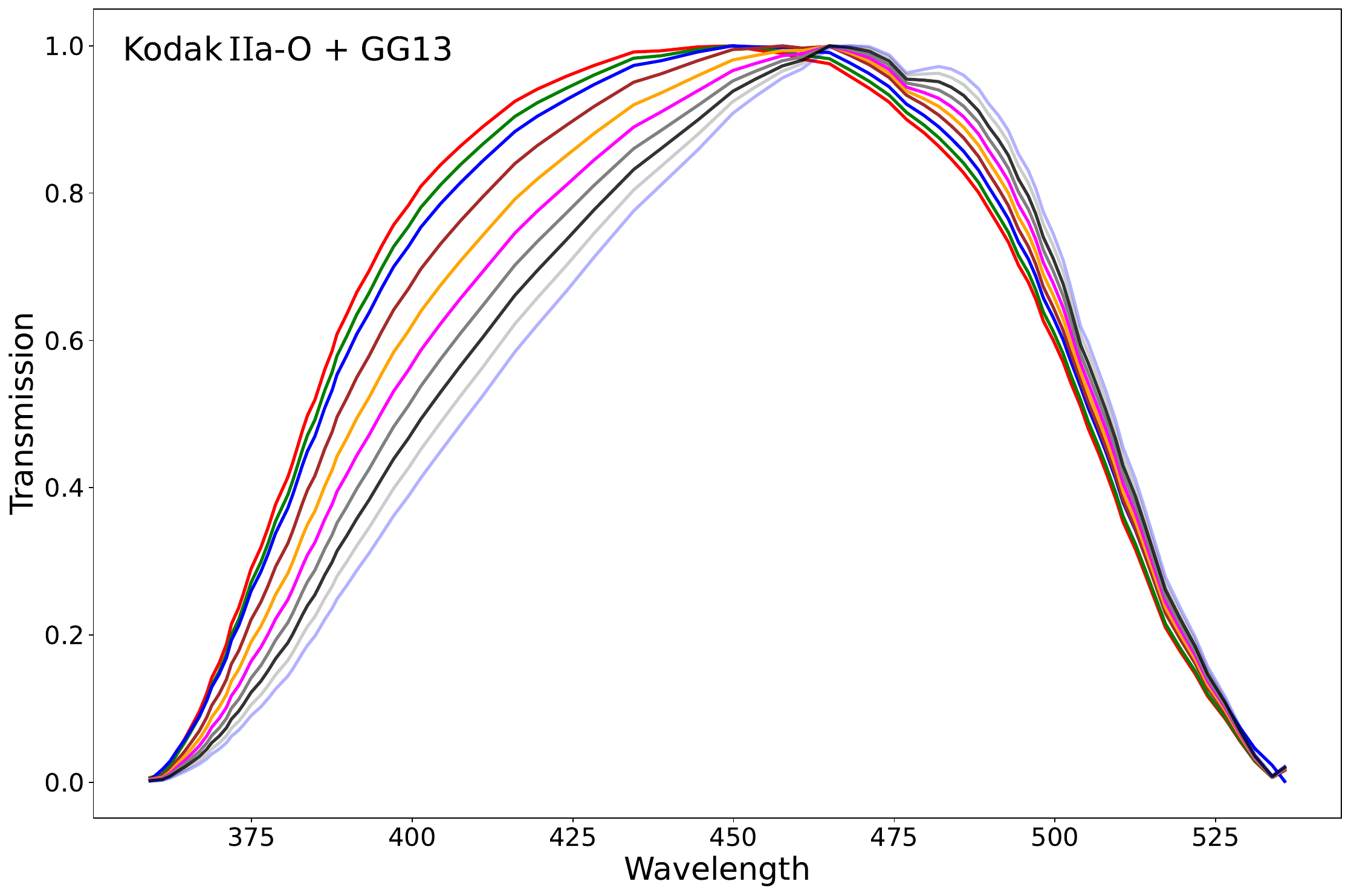}
	\includegraphics[width=0.495\hsize]{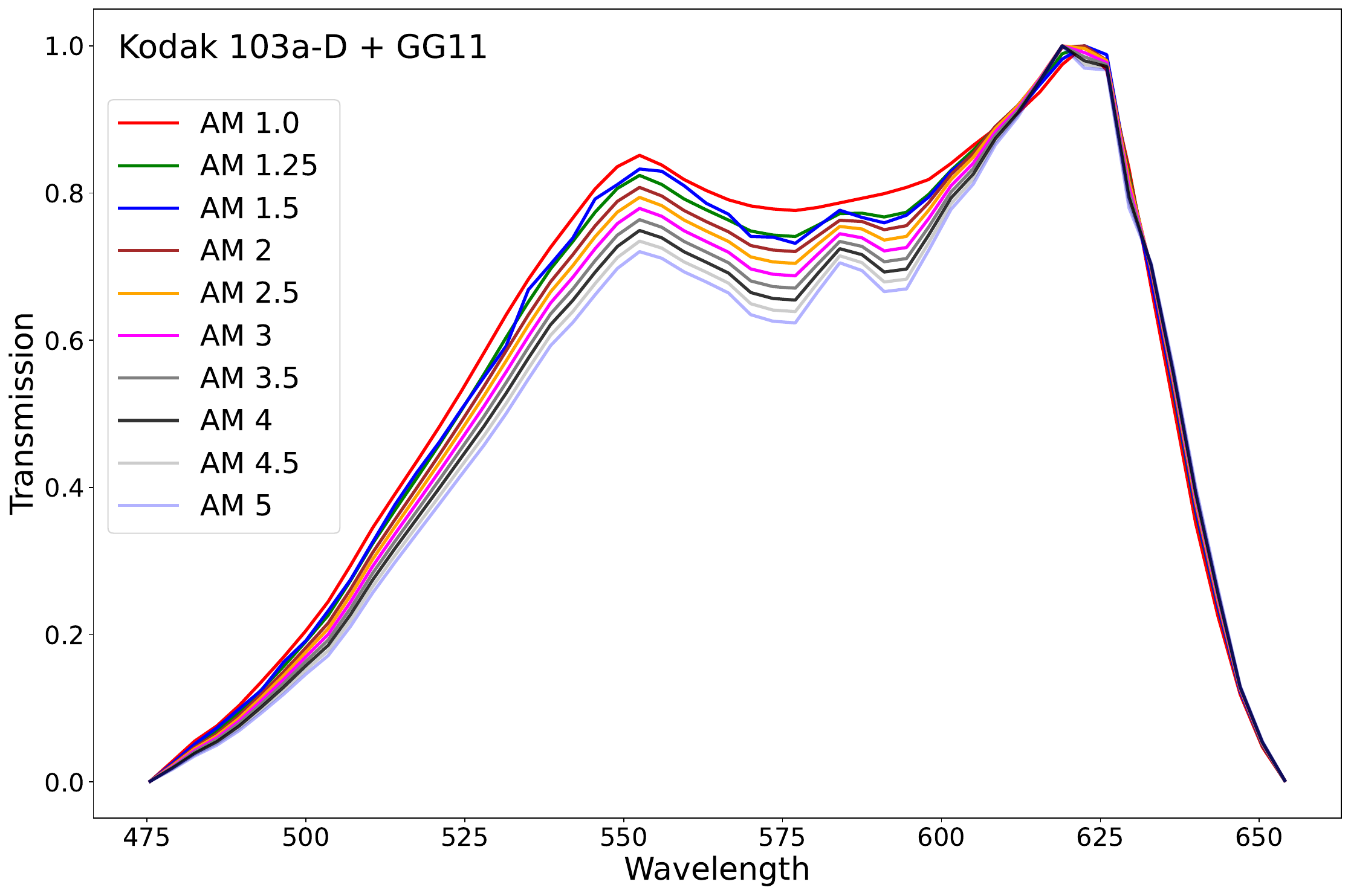}     
	
	\caption{Spectral characteristic curves of Kodak emulsions with Schott filter combinations for photographic photometry in different air masses. Left: Kodak IIa-O emulsion sensitivity curve behind GG13(=385)  filter. Right: Kodak 103a-D emulsion sensitivity curve with GG11 filter.}
	\label{figC1}
\end{figure*}

\end{appendix}

\end{document}